\documentclass[fleqn,usenatbib]{mnras}

\usepackage{newtxtext,newtxmath}
\usepackage[T1]{fontenc}
\usepackage{ae,aecompl}

\usepackage{graphicx}	% Including figure files
\usepackage{amsmath}	% Advanced maths commands
\usepackage{amssymb}	% Extra maths symbols
\usepackage{multirow}
\usepackage[flushleft]{threeparttable}

\title[The local dark matter density]{Measuring the local dark matter density with LAMOST DR5 and Gaia DR2}

\author[Guo et al.]{
Rui Guo$^{1,2}$\thanks{E-mail: guorui13@bao.ac.cn},
Chao Liu$^{1}$,
Shude Mao$^{3,1}$,
Xiang-Xiang Xue$^{4}$,
R.J. Long$^{3,1,5}$,
Lan Zhang$^{1}$
\\
$^{1}$National Astronomical Observatories, Chinese Academy of Sciences, 20A Datun Road, Chaoyang District, Beijing 100101, China\\
$^{2}$University of Chinese Academy of Sciences, Beijing 100049, China\\
$^{3}$Department of Astronomy and Tsinghua Centre for Astrophysics, Tsinghua University, Beijing 100084, China\\
$^{4}$CAS Key Laboratory of Optical Astronomy, National Astronomical Observatories, Chinese Academy of Sciences, Beijing 100101, China\\
$^{5}$Jodrell Bank Centre for Astrophysics, Department of Physics and Astronomy, The University of Manchester, Oxford Road, Manchester M13 9PL, UK\\
}

\date{Accepted XXX. Received YYY; in original form ZZZ}

\pubyear{20XX}

\begin{document}
\label{firstpage}
\pagerange{\pageref{firstpage}--\pageref{lastpage}}
\maketitle

\begin{abstract}
We apply the vertical Jeans equation to the kinematics of Milky Way stars in the solar neighbourhood to measure the local dark matter density. More than 90,000 G- and K-type dwarf stars are selected from the cross-matched sample of LAMOST DR5 and Gaia DR2 for our analyses. The mass models applied consist of a single exponential stellar disc, a razor thin gas disc and a constant dark matter density. We first consider the simplified vertical Jeans equation which ignores the tilt term and assumes a flat rotation curve. Under a Gaussian prior on the total stellar surface density, the local dark matter density inferred from Markov Chain Monte Carlo simulations is $0.0133_{-0.0022}^{+0.0024}\ {\rm M}_{\odot}\,{\rm pc}^{-3}$. The local dark matter densities for subsamples in an azimuthal angle range of $-10^{\circ} < \phi < 5^{\circ}$ are consistent within their 1$\sigma$ errors. However, the northern and southern subsamples show a large discrepancy due to plateaux in the northern and southern vertical velocity dispersion profiles. These plateaux may be the cause of the different estimates of the dark matter density between the north and south. Taking the tilt term into account has little effect on the parameter estimations and does not explain the north and south asymmetry. Taking half of the difference of $\sigma_{z}$ profiles as unknown systematic errors, we then obtain consistent measurements for the northern and southern subsamples. We discuss the influence of the vertical data range, the scale height of the tracer population, the vertical distribution of stars and the sample size on the uncertainty of the determination of the local dark matter density.
\end{abstract}


%%%%%%%%%%%%%%%%%%%%%%%%%%%%%%%%%%%%%%%%%%%%%%%%%%

%%%%%%%%%%%%%%%%% BODY OF PAPER %%%%%%%%%%%%%%%%%%

%%%%%%%%%%%%%%
\section{Introduction}
\label{sec_intro}
The local dark matter density is an important parameter for deriving the Milky Way's overall density profile. It is a local normalisation for different dark matter density profiles. Comparisons of its estimates with the Galactic rotation curve help to constrain the shape of the dark matter halo \citep{sofue2009, weber2010}. This is helpful in understanding the merger history of Milky Way \citep[e.g.][]{read2008, read2009} and galaxy formation models \citep[e.g.][]{dubinski1994, ibata2001, kazantzidis2004, maccio2007, debattista2008, lux2012}. In addition, the local dark matter density is an essential quantity for predicting signals in direct detection experiments of dark matter particles \citep[e.g.][]{lewin1996, frandsen2012, bhattacharjee2013, fairbairn2013, green2017}. The collision rate between the detector material and dark matter particles is proportional to the local dark matter density. An independent measurement helps to break the degeneracy between this density, the mass of the dark matter particles and the scattering cross section \citep{lewin1996, peter2011}.

Since the pioneering work of \cite{oort1932}, many studies have tried to measure the local dark matter density. Different modelling methods and observational samples have been utilised and different results are obtained. In general, the approaches used in these works can be separated into two types \citep{read2014}. The first one extrapolates the local dark matter density from the Milky Way's rotation curve derived from the compilation of kinematic measurements of gas, stars and masers \citep[e.g.][]{merrifield1992, dehnen1998, sofue2009, catena2010, weber2010, mcmillan2011, piffl2014, mcgaugh2016, benito2019, eilers2019, karukes2019}. This method is usually based on a parameterized dark matter density profile and sometimes on priors for some parameters of the baryonic mass models (such as the total stellar surface density $\Sigma_{\star}$, the disc scale length $R_{\rm d}$, the disc scale height $z_{\rm h}$). The second method derives the local dark matter density by analysing the kinematics of stars in the solar neighbourhood using the vertical Jeans equation or the distribution function \citep[e.g.][]{kuijken1989a, kuijken1989b, kuijken1989c, holmberg2000, holmberg2004, garbari2012, zhang2013, bienayme2014, xia2016, hagen2018, sivertsson2018, buch2019, widmark2019}. Similarly, a prior on the total stellar surface density or the local stellar volume density from stellar census is usually used to reduce the degeneracy between baryon components and dark matter.

For approaches using local tracers, there are significant differences in the modelling methods, the assumptions and simplifications made for the modelling, and the observational data used. These ingredients lead to different results and error estimations. In order to match the observations with the models, several ways are used. One way is to integrate the vertical force $K_{z}$ to derive the model velocity dispersion profile, and then compare with the observations \citep[e.g.][whose approach we adopt here]{xia2016}. Another way is integrating $K_{z}$ from $z$ to infinity whilst the baryonic components below $z$ are drawn from the baryon census \citep[e.g.][]{sivertsson2018}. A third way is to calculate a one-dimensional distribution function of the vertical energy, and then compare it with the observed phase-space distribution in the distance-velocity plane \citep[e.g.][]{kuijken1989a, kuijken1989b}. A fourth way is to compare the observed and model predicted number density profiles, with the latter being derived from the observed vertical velocity distribution function and an assumed potential \citep[e.g.][]{holmberg2000, holmberg2004}.

Model comparison is complicated by differences and uncertainties in the observed data. Nevertheless, \cite{garbari2011} tried to compare their so-called minimum assumption method with the Holmberg and Flynn method \citep{holmberg2000} using an N-body simulation. They found that the methods, which assume that the distribution function of a tracer population is only a function of the vertical energy, become systematically biased when the motion of the tracers is not truly separable in $z$. This effect becomes significant when the data extends beyond one disc scale height ($\sim 300$ pc). See the review article \cite{read2014} for more detailed comparisons of previous works.

Besides the different methods, the data used in previous works vary greatly in the sample size, the sky coverage, the type of tracer and the accuracy of the stellar parameters. With development in astronomical techniques, several new measurements came out from new Galactic surveys, such as the  Sloan  Digital  Sky  Survey \citep[hereafter SDSS;][]{smith2012, zhang2013}, the RAdial Velocity Experiment \citep[hereafter RAVE;][]{siebert2008}, the Large Sky Area Multi-Object Fibre Spectroscopic Telescope (hereafter LAMOST) survey \citep[e.g.][]{xia2016} and the Gaia satellite \citep[e.g][]{buch2019, widmark2019}. There are also works combining data from different surveys, such as \cite{hagen2018} who combined data from TGAS (Tycho-Gaia Astrometric Solution) and RAVE. These surveys cover different areas, and have different advantages and disadvantages.

The tracers used vary from K-dwarf stars \citep[e.g.][]{kuijken1989b}, red clump stars \citep[e.g.][]{bienayme2014, hagen2018}, G-dwarfs separated into $\alpha$-young and $\alpha$-old populations \citep[e.g.][]{budenbender2015, sivertsson2018} and stars divided into eight samples within the G-band magnitude range of 3.0 to 6.3 \citep{widmark2019}. Different tracers yield quite different results \citep[e.g.][]{sivertsson2018, buch2019, widmark2019}.

A possible explanation of the different results for different tracers is that the stellar disc is not in dynamical equilibrium in the solar neighbourhood. From different surveys, there is substantive observational evidence of vertical oscillations of the stellar disc, causing it to act as a ringing, wobbling or flaring disc \citep[e.g.][]{widrow2012, williams2013, xu2015, carrillo2018, wang2018, bennett2019, wang2019, gardner2020}. These may have been caused by the recent passage of a massive satellite such as the Sagittarius dwarf galaxy \citep[e.g.][]{purcell2011, widrow2014, onghia2016}, or by disturbance from the spiral arms \citep[e.g.][]{antoja2011, faure2014, monari2016}. Disequilibrium of the disc can thus make traditional Jeans modelling of the local dark matter density problematic \citep[e.g.][]{widrow2012, read2014, haines2019}. In combination with other systematics, this could result in different determinations of the local dark matter density.

Utilising the vertical Jeans equation used in \cite{xia2016}, this work combines the LAMOST fifth data release (DR5) and the Gaia second data release (DR2) to select a well-defined data sample. Selection effects, volume completeness, accuracy of the distance measurements, proper motion and line of sight velocity measurements are carefully considered. \cite{xia2016} selected 1427 stars from LAMOST DR2 with galactic latitude $b\ > 85^{\circ}$ together with some other criteria. The galactic latitude criterion is used to guarantee that the radial velocities are approximately equal to the vertical velocities. This is necessary due to the lack of traverse motions. With proper motions from Gaia, we can select a sample in a column, which can cover a larger azimuthal angle range and have a considerably larger sample size ($\sim 65$ times). These help us obtain more reliable estimates, and enable us to compare the dark matter densities measured within different regions.

The paper is organized as follows. In Section \ref{sec_data}, we describe the selection criteria of our sample and how the selection effects are corrected. In Section \ref{sec_method}, we present the assumptions, simplifications and mass models applied in the vertical Jeans equation, and how we estimate the parameters through the Markov Chain Monte Carlo (hereafter MCMC) technique. The results under different priors are shown in Section \ref{sec_result}. In Section \ref{sec_discuss}, we present some comparisons, and discuss the asymmetry between the Galactic north and south. Some mock tests for the systematic uncertainties are also presented in this section. Finally, in Section \ref{sec_conclu}, we present our conclusions. Throughout the paper, we adopt a solar position of ($-$8.34, 0., 0.027) kpc in the Galactic Cartesian coordinates system \citep{reid2014b, chen1999}, and a solar peculiar velocity, relative to the local standard of rest, of (9.58, 10.52, 7.01) km s$^{-1}$ in the radial, azimuthal and vertical directions, respectively \citep{tian2015}.

%%%%%%%%%%%%%%
\section{Data}
\label{sec_data}
%%%%%%%%%
\subsection{Selection criterion}
\label{ssec_dsc}
LAMOST, also known as the Guo Shou Jing Telescope, is a 4 metre reflective Schmidt telescope with 4000 fibres in a field of view of 20 deg$^2$ in the sky \citep{cuixq2012, zhaog2012}. The design makes it the most efficient spectroscopic survey telescope for bright stars in the Milky Way. The survey provides millions of stellar spectra, which can be used to study the structure, formation and evolution of Milky Way \citep{denlc2012}. LAMOST started its pilot survey in 2011 and finished the first-five-year regular survey in 2017. LAMOST DR5, including data from both surveys, contains 9,017,844 low-resolution (R $\sim$ 1,800) spectra in the optical band (3690-9100 \AA~), of which 8,171,443 are stellar spectra.

The Gaia satellite was launched in December 2013 by the European Space Agency. It is designed to provide accurate astrometric and photometric information for billions of sources over the full sky, aiming to produce a three-dimensional map of most of the Milky Way \citep{gaia2016, gaia2018a, gaia2018b}. Gaia DR2 provides five-parameter astrometric measurements (positions, parallaxes, and proper motions) for about 1.3 billion sources \citep{gaia2018a}. The typical uncertainties for sources with a broad-band magnitude G $<$ 15 are between 0.02 and 0.04 milli-arcseconds (mas) for the parallax and 0.07 mas yr$^{-1}$ for the proper motion. These values become larger, to 0.7 mas and 1.2 mas yr$^{-1}$ at G=20. 

The cross-matched sample of LAMOST DR5 and Gaia DR2 contains 8,852,848 common objects. For these objects, we usually have spectroscopic parameters from LAMOST DR5 (e.g. effective temparature $T_{\rm eff}$, surface gravity $\log g$, metallicity [Fe/H], radial velocity V$_{los}$), and astrometric and photometric parameters from Gaia DR2 (e.g. parallax $\varpi$, two proper motions $\mu_{\alpha\star}$ and $\mu_{\delta}$ in equatorial coordinates and Gaia G band apparent magnitude G). With these parameters, we can select a column volume complete G/K dwarf sample with accurate distances, radial velocities, proper motions, and greatly improve the sample size compared to previous works. Distance is estimated from the parallax using a Bayesian inference method following \cite{bailer-jones2018}.

Our sample contains 93,609 G/K-type dwarf stars. This sample is about 65 times larger than the sample selected from the cone volume in \cite{xia2016}, which contained 1427 stars. The selection criteria are as follows (symbols are explained below):
\\

(i) $K_{mag\_2MASS} < 14.3$;

(ii) $\left( \frac{\rm S}{\rm N} \right)_{g} > 20$;

(iii) $\log\,g$ > 4;

(iv) $5000 < T_{\rm eff} < 6000$ K;

(v) self-crossmatch;

(vi) $4.0 < M_{G} < 5.0$;

(vii) $|Z| < 1.3$ kpc;

(viii) distance > 0.2 kpc;

(ix) $|R - R_{\odot}| < 0.2 $ kpc \& $|\phi| < 5^{\circ}$;

(x) $[Fe/H] > -0.4$;

(xi) $\varpi > 0$ \& $\sigma_{\varpi}/\varpi < 0.2$.
%\begin{enumerate}%[leftmargin= 16pt ;i)]
%\item $K_{mag\_2MASS} < 14.3$;
%\item $snrg$ > 20;
%\item $logg$ > 4;
%\item $5000 < T_{\rm eff} < 6000$ K;
%\item self-crossmatch;
%\item $4.0 < M_{G} < 5.0$;
%\item $[Fe/H] > -0.4$;
%\item $|R - R_{\odot}| < 0.2 $ kpc \& $|\phi| < 5^{\circ}$;
%\item $|Z| < 1.3$ kpc;
%\item $\varpi > 0$ \& $\sigma_{\varpi}/\varpi < 0.2$ .
%\end{enumerate}
\\

In the first criterion (i), $K_{mag\_2MASS}$ is the K band magnitude from the Two Micron All Sky Survey \citep[hereafter 2MASS;][]{skrutskie2006}. This criterion is adopted because this magnitude is the limiting magnitude of the 2MASS catalog, which is utilized for the selection effects correction later. The signal-to-noise (S/N) criterion (ii) is adopted to ensure the accuracy of stellar radial velocities. The third criterion (iii) is applied to exclude giant stars using the logarithmic stellar surface density ($\log\,g$). In order to select a tracer population with a specific spectral type, we adopt an effective temperature cut $5000 < T_{\rm eff} < 6000$ K (iv). This cut is a little tighter than that used in \cite{xia2016}. Stars with $T_{\rm eff} > 6000$ K are not selected because they may be dominated by young stars with ages $<$ 4 Gyr, which may not be sufficiently relaxed to be in dynamical equilibrium \citep{tian2015}. The lower $T_{\rm eff}$ threshold can make sure the measurements of the stellar atmospheric parameters are more accurate. After applying these four broad criteria, we perform the self-crossmatch (v) to exclude stars with repeated observations.

\begin{figure}
   \centering
   \includegraphics[width=0.8\columnwidth]{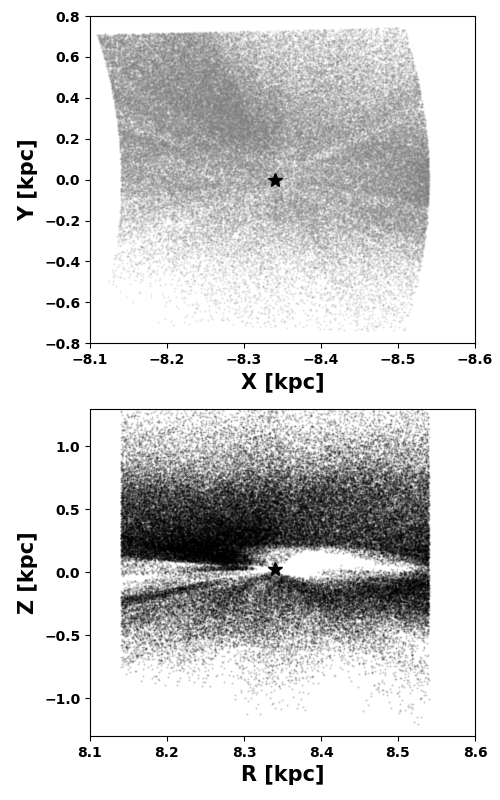}
   \caption{Stellar positions (grey dots) of the sample used in this work in galactocentric Cartesian coordinates. The upper panel is the `X-Y' plane (face-on) projection, while the bottom panel is the `R-Z' plane (edge-on) projection. The positive X direction is towards the Galactic centre, and the positive Z is towards the north Galactic pole. The stars indicate the solar positions. The void on the `R-Z' plane in the galactic anti-centre direction is due to the selection criteria.}
   \label{fig_spos}
\end{figure}

Volume completeness is influenced by the distance and the magnitude range of observed stars. In order to obtain a volume complete sample, we apply the sixth (vi) and the seventh (vii) criteria. The absolute G band magnitude M$_{G}$, is calculated from the G magnitude and the parallax. Stars with distances smaller than 200 pc are excluded to avoid the selection effect in the bright end (viii). $R_{\odot}$ in the criterion (ix) is the solar distance to the Galactic centre, and $\phi$ is the galactic azimuthal angle. The ninth criterion is the volume cut for an annulus with a galactic radius width of 0.4 kpc and an azimuthal angle width of 10$^{\circ}$ ($\sim$ 1.4 kpc). The azimuthal width may be somewhat large, and thus will average the dark matter density azimuthally. This will be discussed later in Section \ref{ssec_rhophi}. Ideally, we would like a single tracer population and would like to use $\alpha$-element abundances to remove thick components from our tracer population. However, we do not have $\alpha$-element abundances for the dwarf stars. Thus we simply utilize a metallicity cut $[Fe/H] > -0.4$ as the tenth criterion (x). This cut effectively removes the thick component from the tracer. The resultant number density profile of the tracer is well-fitted by a single exponential function. Finally, we apply cuts on the parallax and parallax error (xi). The parallax $\varpi$ is required to be positive and the parallax error $\sigma_{\varpi}$ is required to be smaller than 0.2 relative to the parallax. The stellar positions of our selected sample are shown in Fig. \ref{fig_spos}. The void on the `R-Z' plane in the galactic anti-centre direction is due to selection criterion (viii) and the footprint of LAMOST DR5. LAMOST DR5 lacks observations in the region of galactic longitude $\sim 90^{\circ} < l < 160^{\circ}$ and latitude $\sim 0^{\circ} < b < 60^{\circ}$.

%%%%% figure of selection
%%%%%

%%%%%%%%%
\subsection{Selection effect correction}
\label{ssec_dse}
In general, the observed number density is different from the real number density of a tracer population due to selection effects. The selection function of a spectroscopic survey is affected by the survey's targeting strategy, the actual observational conditions, data reduction and so on. Thus, we need to correct the selection function for our sample to derive the real number density.

For the LAMOST survey, the simple targeting algorithm designed by \cite{carlin2012} has not been fully used for a few reasons. Consequently, the LAMOST survey separates the targets into different plates with different ranges of magnitudes in each line-of-sight. In addition, at least four different catalogs based on different photometric systems are adopted as the source catalogs for targeting \citep[see][for more details]{liu2017}. To avoid complicated calibrations, we finally choose the 2MASS catalog \citep{skrutskie2006} as the complete photometric dataset, which covers most of the LAMOST observed stars.

Our sample selection effects are corrected in three steps. Firstly, for each observed plate of LAMOST DR5, we count the numbers of stars from the photometric ($N_{\rm photo}$) and spectroscopic ($N_{\rm spec}$) samples in the same $J - K$ vs. $K$ grid. This colour-magnitude (CM) map has enough large grid ranges and sufficiently small grid spaces. Then we can assign a weight by dividing $N_{\rm photo}$ by $N_{\rm spec}$ to each pixel of the CM map, and thus each star has a weight according to its position in the CM map. Secondly, for each plate, i.e. each line-of-sight, we choose the same distance grids. We calculate the probility of a star being in each distance grid according to the star's distance and distance errors. Then we sum up all the probilities for each grid by multiplying the weight calculated in the first step. Dividing by the solid angle of the plate and the distance square of the grid, we can derive the number density of each grid. Thus the number density values can be obtained by linearly interpolating the number density function for all stars in that plate. Finally, all stars in our sample are separated into different plates and their number density values are calculated by repeating the previous two steps.

\begin{figure}
   \centering
   \includegraphics[width=\columnwidth]{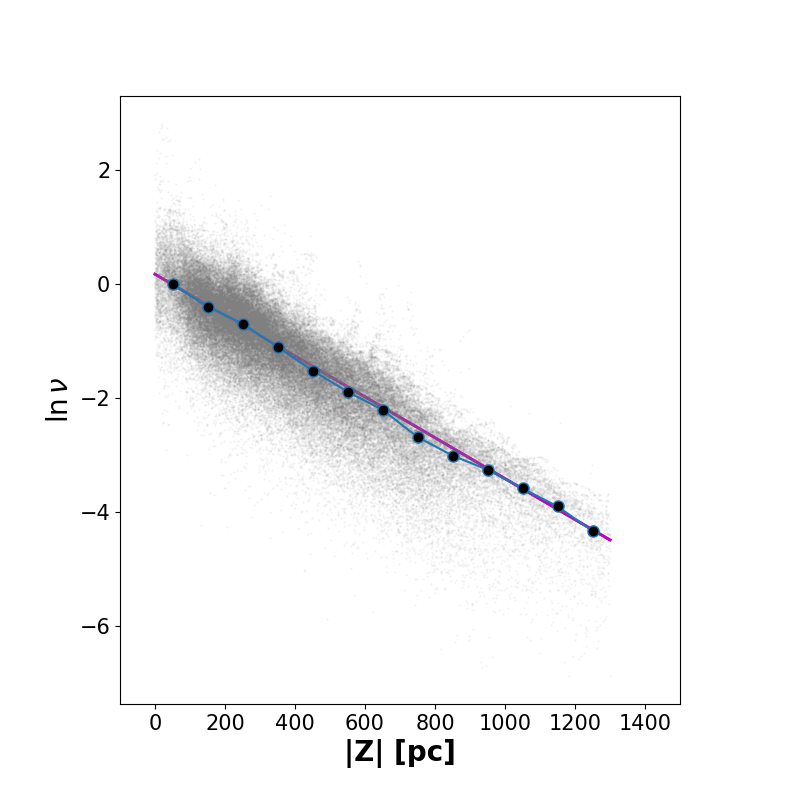}
   \caption{The number density profile of the selected sample. The grey dots are the number densities for all stars with the selection effects corrected. The black dots with error bars (from bootstrapping) show the binned number density profile of grey dots with a bin size of 100 pc. The magenta line is an exponential fit to the dots. All the number densities are normalized to the first bin.}
   \label{fig_nu}
\end{figure}

The number density profile of our observational sample, after the selection effects have been corrected, is shown in Fig. \ref{fig_nu}. The number density is binned with a bin size of 100 pc, and errors are obtained through bootstrapping. The binned number density can be well fitted with a single exponential function, shown as the magenta line in Fig. \ref{fig_nu}. It implies that the majority of stars in our sample belong to the thin disc with a single scale height ($\rm h_{1}$) of $278.6 \pm 3.7$ pc. This scale height will be used in the vertical Jeans equation in the next section.

The 3D velocities are calculated with the estimated distance, two position parameters, two proper motions and the radial velocity. A 240 ${\rm km\,s^{-1}}$ circular motion of the LSR \citep{piffl2014} is taken to transform the velocities into Galactic rest frame velocities. Gaia DR2 also provides radial velocities for 7.2 million sources. These radial velocities are the median values averaged over the 22 month time span of the observations. Their uncertainties show dependence on the stellar effective temperature and the magnitude in the $G_{RVS}$ photometric band, where the values are approximately 1.4 km s$^{-1}$  at $G_{\rm RVS} = 11.75$ for stars with $T_{\rm eff} \sim 5500$ K \citep{gaia2018a}. About 26,000 stars in the our sample have radial velocity measurements from Gaia. However, the vertical heights of these stars are smaller than 600 pc, which is too small for our method. Note that LAMOST radial velocities have a systematic offset $\sim 5.4$ ${\rm km\,s^{-1}}$ \citep{tian2015} compared to the APOGEE data \citep{ahn2014}. The reason for this offset is not known. This LAMOST systematic offset is $\sim 5.3$ ${\rm km\,s^{-1}}$ when the comparison is with Gaia radial velocities. This systematic offset has been corrected in our samples.

\begin{figure}
   \centering
   \includegraphics[width=\columnwidth]{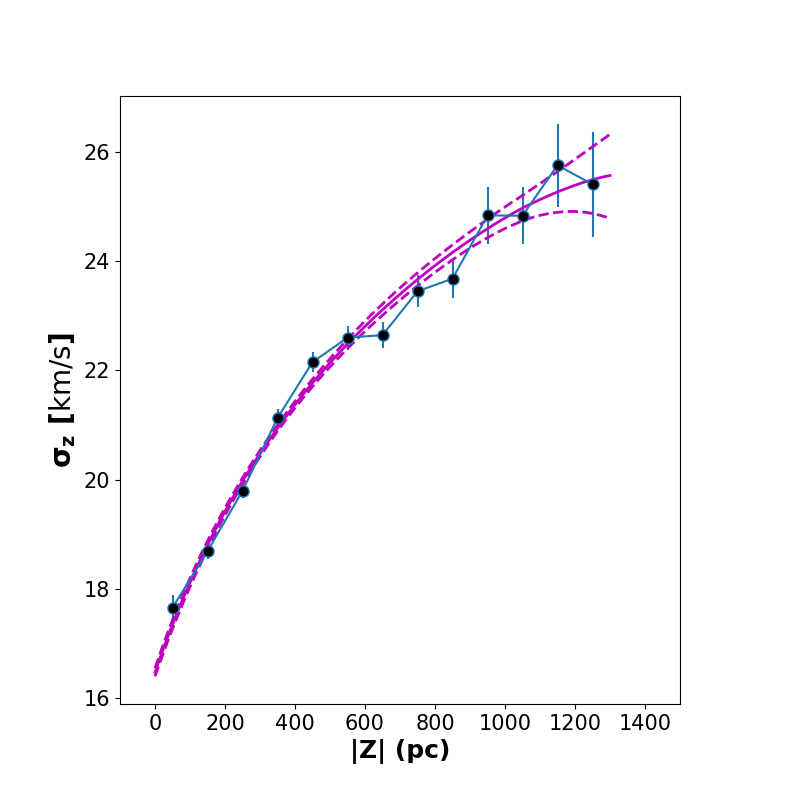}
   \caption{The vertical velocity disperion profile of our observational sample. The black dots with error bars indicate the binned vertical velocity dispersion profile of our sample. The magenta solid line is the median of profiles calculated from the selected parameters in the MCMC program, while the two magenta dashed lines show the 1$\sigma$ region for those profiles.}
   \label{fig_sigmav}
\end{figure}

The vertical velocity dispersion profile of our sample is shown as dots in Fig. \ref{fig_sigmav}. The velocity dispersions are calculated from the standard deviations of the vertical velocities, while the measurement errors are removed by subtracting a systematic instrumental error of 4.5 ${\rm km\,s^{-1}}$ \citep{gao2014}. Error bars are estimated using bootstrapping. Note that binning velocity dispersion is just used for plotting. We utilize the spatial and kinematic information for individual stars without binning for estimating our model parameters (see the Section \ref{ssec_mcmc}).

%%%%% figure of number density
%%%%%

%%%%% figure of velocity dispersion
%%%%%

%%%%%%%%%%%%%%
\section{Method}
\label{sec_method}
%%%%%%%%%%%%%%
\subsection{The vertical Jeans equation}
\label{ssec_jeans}
Following \cite{zhang2013} and \cite{xia2016}, we use the vertical Jeans equation method to measure the local dark matter density. The first two assumptions we make are as follows.
\\

(1) The system is in steady state.

(2) The gravitational potential of the Milky Way is axisymmetric.
\\

With these two assumptions, we can integrate the collisionless Boltzmann equation over the vertical momentum to derive the vertical Jeans equation in cylindrical coordinates:
%%%%%%
\begin{equation}
\frac{\partial}{\partial z}(\nu\sigma^{2}_{zz})\ +\ \frac{1}{R}\frac{\partial}{\partial R}(R\nu\sigma^{2}_{Rz}) = -\nu\frac{\partial \Phi}{\partial z}\ ,
\label{eq_full_jeans}
\end{equation}
%%%%%%
where $\sigma_{ij}^{2} = \overline{\upsilon_{i}\upsilon_{j}}\ -\ \overline{\upsilon_{i}}\ \overline{\upsilon_{j}}$ is the velocity dispersion tensor \citep{binney2008}, and $\nu$ is the vertical number density of the tracer population moving in the Galactic potential $\Phi$. The mean vertical velocity $\overline{\upsilon_{z}}$ is assumed to be zero here. The vertical velocity dispersion $\sigma_{z}^{2}$, $\nu$ and $\Phi$ are only functions of the vertical height $z$.

The Galactic potential $\Phi$ is connected with the mass distribution by the Poisson equation. Integrating the Poisson equation, we obtain the function of the gravitational force perpendicular to the Galactic plane, i.e. the $K_{z}$ force. To simplify the equation and get the model predicted velocity dispersion, we utilize the following assumptions.
\\

(3) the tilt term $\frac{1}{R}\frac{\partial}{\partial R}(R\nu\sigma^{2}_{Rz})$ is ignored initially for simplicity. The tilt term couples the radial and vertical motions. Following the discussions of the asymmetric drift in \cite{binney2008}, the tilt term is smaller than the first term in Eq. \ref{eq_full_jeans} at least by a factor of $2zz_{\rm h}/(\rm R_{\rm d}R_{\odot})$, as also discussed in \cite{garbari2011} and \cite{zhang2013}. Here $z_{\rm h}$ and ${\rm R}_{\rm d}$ are the scale height and the scale length of the disc, respectively. For a volume with a small width of 0.4 kpc in R and a vertical range of 1.3 kpc, this factor is about 0.03. Thus there is a good reason for ignoring the tilt term in the simplified model, but we reconsider it in Section \ref{sssec_NS_tilt} as a possible solution for solving the velocity asymmetry between the Galactic north and south.

(4) the rotation curve is flat with both R and $z$ in the solar neighbourhood \citep{binney2008, bovy2012}, i.e. the contribution of the circular velocity term is negligible. The contribution of the circular velocity term can be quantified via the Oort constants A and B \citep{binney2008}:
%%%%%%%%%
\begin{equation}
\frac{1}{4\pi G R} \frac{\partial V_{c}^{2} (R,z)}{\partial R} = \frac{B^{2} - A^{2}}{2\pi G}\ .
\label{eq_vcterm}
\end{equation}
%%%%%%%%%
We need to add this term to our estimated dark matter density. Usually, $B^{2} < A^{2}$, which implies that neglecting the circular velocity term will overestimate the dark matter density. According to different measurements \citep[e.g.][]{gunn1979, feast1998, fernandez2001, olling2003}, this term has a contribution about $0 - 0.003$ ${\rm M}_{\odot}\,{\rm pc}^{-3}$ to the local dark matter density. Nevertheless, as long as the circular velocity term is independent on the vertical height $z$, this term can be individually estimated and the estimated dark matter density can be simply corrected. In this work, we always ignore this term.

(5) the total mass density consists of stars, gas and dark matter.

(6) the gas disc is razor thin without thickness, and the total gas surface density ($\Sigma_{\rm gas}$) is 13.2 ${\rm M}_{\odot}\,{\rm pc}^{-2}$ \citep{flynn2006}. Further discussion about the gas model is in Section \ref{ssec_gas}.

(7) the dark matter density is constant ($\rho_{\rm dm}$).

(8) the stellar disc is a single exponential disc with a small scale height $z_{\rm h}$, i.e. the thick disc is negligible. Our rationale for so doing is as follows. The thick disc is estimated to be a small fraction of the stellar midplane density (about 10\%) with a scale height of about 1200 pc \citep{flynn2006, juric2008}. The surface density of the thick disc at $z= 1300$ pc is then $\sim$ 7 ${\rm M}_{\odot}\,{\rm pc}^{-2}$, which is about 10\% relative to the total surface density. Taking a local dark matter density of $0.01\, {\rm M}_{\odot}\,{\rm pc}^{-3}$, ignoring the thick disc will lead to a maximum uncertainty of 26\% to the local dark matter estimation. We revisit our rationale in Section \ref{ssec_double} where we also consider a double disc model with a thick disc.\\

With these assumptions, the $K_{z}$ function can be expressed as:
%%%%%%%
\begin{align}
K_{z}(z) & \equiv -\ \frac{{\rm d} \Phi}{{\rm d} z} = -\int_{0}^{z} 4\pi{\rm G}\rho_{\rm tot}(z')\, dz' = -2\pi G \Sigma_{\rm tot}(z) \nonumber\\
&=\ -2\pi G \left\{ \Sigma_{\star} \left[ 1 - \exp\left( -\frac{z}{z_{\rm h}} \right) \right] + \Sigma_{\rm gas} + 2 \rho_{\rm dm}z \right\}\ ,
\label{eq_kz_int}
\end{align}
where $\rho_{\rm tot}(z)$ and $\Sigma_{\rm tot}(z)$ are the total mass density and total surface mass density up to a vertical height $z$, respectively. The latter is connected with the $K_{z}$ force by $\Sigma_{\rm tot}(z) = K_{z}/(-2 \pi G)$. $\Sigma_{\star}$ is the total stellar surface density and $\rho_{\rm dm}$ is the constant dark matter density we seek in this work.

%%%%%%%%%%%%%
\subsection{Parameter estimation with MCMC}
\label{ssec_mcmc}
In observations, we have the number density profile of the chosen tracer population and the vertical velocities of stars. The former is assumed to be a single exponential profile,
%%%%%%%%%%%
\begin{equation}
\nu(z) = \nu_{0} \exp \left( -\frac{z}{h_{1}} \right)\ ,
\label{eq_nuz}
\end{equation}
%%%%%%%%%%%
which is a quite good approximation to our data as shown in Fig. \ref{fig_nu}. The vertical velocities can be compared to the model velocity dispersion profile derived from the vertical Jeans equation. Inserting the $K_{z}$ function into Eq. \ref{eq_full_jeans} and integrating this equation on both sides, we can obtain:
%%%%%%%%%%%
\begin{align}
& \nu(z) \sigma_{z}^{2}(z)\ -\ \nu(z_{0}) \sigma_{z}^{2}(z_{0})\ =\ 
\int_{z_{0}}^{z} \nu(z') K_{z}(z') dz' , \nonumber\\
&\sigma_{z}^{2}(z)\ =\ f(z)\ +\ \frac{\nu(z_{0}) \sigma_{z}^{2}(z_{0}) - \nu(z_{0}) f(z_{0})}{\nu_{0} \exp \left( -\frac{z}{h_{1}} \right)}\ ,
\label{eq_sigz}
\end{align}
%%%%%%%%%%%
where
%%%%%%%%%%%
\begin{equation}
  \begin{split}
f(z) =\ & 2\pi G h_{1} \left\{\ \Sigma_{\star} \left[\ 1 - \frac{z_{\rm h}}{h_{1} + z_{\rm h}} \exp\left( -\frac{z}{z_{\rm h}} \right)\ \right] \right. \\
      &\left. +\ \Sigma_{\rm gas}\ +\ 2 \rho_{\rm dm}(z + h_{1})\ \right\}\ ,
  \end{split}
\label{eq_fz}
\end{equation}
%%%%%%%%%%%%
$z_{0}$ is the integration boundary, which can be arbitrary. The contributions of stars, gas and dark matter to the velocity dispersion are different as shown in Eqs \ref{eq_sigz} and \ref{eq_fz}. The contribution of the razor thin gas disc is constant. The contribution from the stellar disc increases as a negative exponential with $z$ and approaches flat beyond about two scale heights ($\geq 1$ kpc). The dark matter, which provides a linearly increasing contribution along $z$, dominates the profile at high-$z$ region. Thus, with data covering over larger vertical range, we can separate the dark matter from the baryonic components more easily.

With equations (\ref{eq_sigz}) and (\ref{eq_fz}), we can compare the model velocity dispersion profile ($\sigma_{z,{\rm model}}(z)$) with the observed stellar vertical velocities ($\upsilon_{z}$). Following \cite{xia2016}, we use the MCMC technique rather than binning the data to obtain estimates of model parameters. That is because the latter will lose spatial information. The MCMC package we use is {\tt EMCEE} \citep{foreman2013}. 

There are three differences in the parameter selection between this work and \cite{xia2016}. First, we leave the boundary condition $\sigma_{z}(z_{0} = 50\,{\rm pc})$ as a free parameter and thus all stars are used in MCMC. In \cite{xia2016}, stars with $z$ between 100 and 300 pc are used as boundary condition and are not taken into the parameter estimation in MCMC. Shown as the second term on the right-hand side of Eq. \ref{eq_sigz}, the boundary condition term provides an exponentially increasing contribution. Thus our treatment will be better especially when the chosen $\sigma_{z}(z_{0})$ has a large uncertainty or biased against the true value.

Second, the number density $\nu(z)$ is individually fitted by binning the data before it is taken into the model velocity dispersion calculation for two reasons. One is that the tracer number density is independent of the mass models and can be determined separately. As shown in Fig. 6 of \cite{xia2016}, the scale height $h_{1}$ is almost uncorrelated with other parameters. The second is that we have more stars in the low-$z$ regime. If we add $\nu(z)$ into MCMC modelling, it will be biased and the scale height will be slightly underestimated.

When comparing stellar velocties $\upsilon_{z}$ with $\sigma_{z,{\rm model}}(z)$, we calculate the probability of $\upsilon_{z}$ in a Gaussian distribution, which has a standard deviation of $\sigma_{z,{\rm model}}(z)$ and a mean velocity $\overline{\upsilon}$. For a disc in perfect equilibrium, $\overline{\upsilon} = 0$. But for the Milky Way, this is not true. The last difference is the treatment with the mean velocity $\overline{\upsilon}$. It is a free parameter with the same value at different vertical heights in \cite{xia2016}. However, this parameter has quite a small influence on the modelling and is not shown in their article. Here, we derive a mean velocity profile with a bin size of 100 pc. For stars in the same bin, we use the same mean velocity.

The remaining parameters in our MCMC modelling are denoted as $\boldsymbol{p} = (\Sigma_{\star},\ z_{\rm h},\ \rho_{\rm dm},\ \sigma_{z}(z_{0}))$. The log-likelihood is given by
%%%%%%%%%%
\begin{align}
{\rm ln}\ L = &-\ \sum_{i} {\rm ln} \left[ \sqrt{2\pi}\sigma_{z,{\rm model}}(z_{i}) \right] \nonumber\\
&-\ \frac{1}{2}\sum_{i} \left[ \frac{\upsilon_{i} - \overline{\upsilon_{i}}}{\sigma_{z,{\rm model}}(z_{i})} \right]^{2}\ ,
\label{eq_logl}
\end{align}
%%%%%%%%%%
where $i$ is the stellar label and $\overline{\upsilon_{i}}$ is the mean velocity from the binned mean velocity profile.

\subsection{Tilt term}
\label{ssec_tilt}
\begin{figure}
   \centering
   \includegraphics[width=0.9\columnwidth]{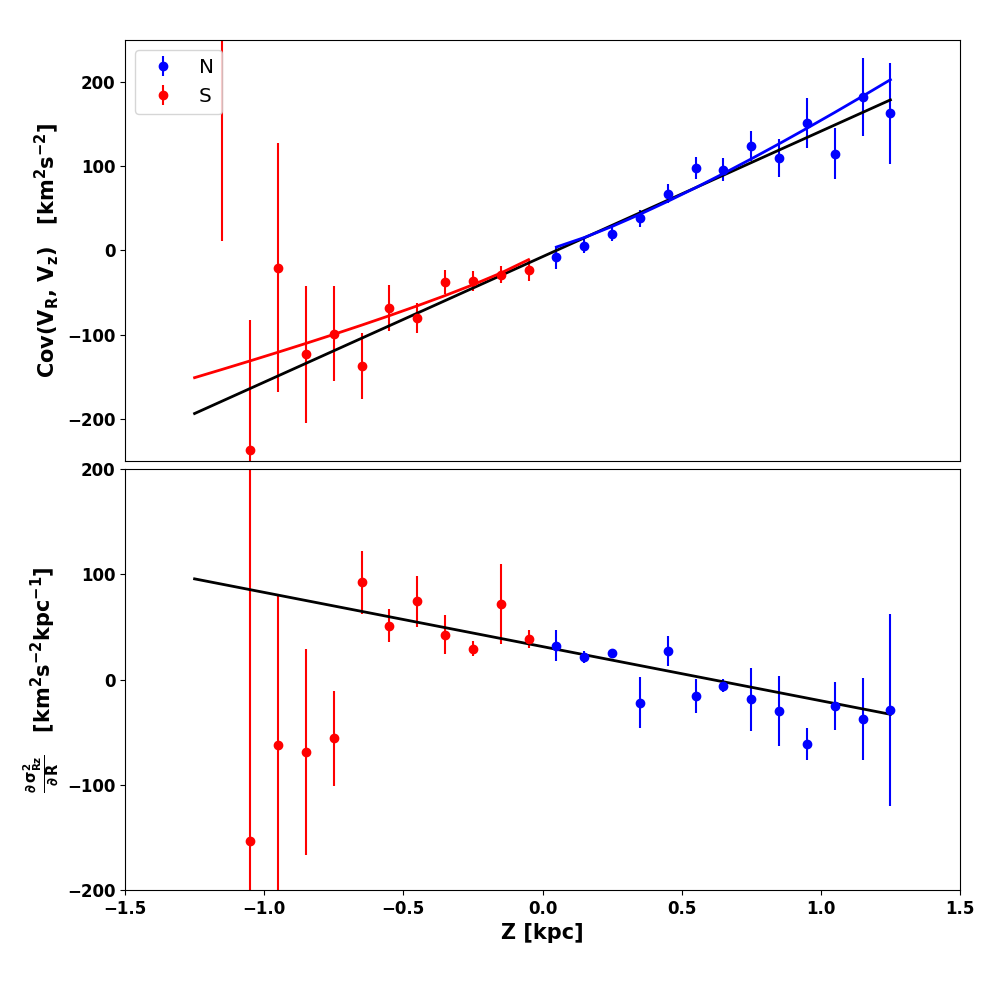}
   \caption{$\sigma_{Rz}^{2}$ profile (upper panel) and its derivative with respect to R (lower panel) for both the northern (blue dots) and southern (red dots) subsamples. The blue and red solid lines in the upper panel are the model fits of the function $Az^{n}$ to the northern and southern $\sigma_{Rz}^{2}$ profiles, respectively. The black lines in both panels are linear fits to the two profiles, which combine the northern and southern subsamples together.}
   \label{fig_NS_sigRz}
\end{figure}

In all analyses except for the Section \ref{sssec_NS_tilt}, the tilt term is ignored. We take the tilt term into consideration in Section \ref{sssec_NS_tilt} in order to check if this term could explain the velocity asymmetry between north and south. The tilt term can be separated into two components:
%%%%%%
\begin{equation}
\frac{1}{R}\frac{\partial}{\partial R}(R\nu\sigma^{2}_{Rz}) = \nu\sigma^{2}_{Rz}(\frac{1}{R}\, -\, \frac{1}{h_{R}})\, +\, \nu\frac{\partial \sigma^{2}_{Rz}}{\partial R} ,
\label{eq_tilt}
\end{equation}
%%%%%%
where $h_{R}$ is the scale length of the tracer population. For the covariance of the radial and vertical velocities ($\sigma^{2}_{Rz}$), we utilise a power law function $Az^{n}$ to model it. The fitting of $\sigma^{2}_{Rz}$ is shown in the upper panel of Fig. \ref{fig_NS_sigRz}. The $\sigma^{2}_{Rz}$ values for the northern and southern subsamples are slightly different. The first term on the right hand of Eq. \ref{eq_tilt} can be analytically obtained after the scale length is set as $h_{R} = 2.5$ kpc. 

To calculate the derivative of $\sigma^{2}_{Rz}$ with respect to R, we choose 4 radial bins, with a bin width of 0.4 kpc. The centres of those four bins are 7.94, 8.34, 8.74, 9.14 (kpc). For each vertical height, we utilise linear fitting to $\sigma^{2}_{Rz}$ to obtain its derivative on R. The derivatives of $\sigma^{2}_{Rz}$ for bins in the northern and southern sky are shown in the lower panel of Fig. \ref{fig_NS_sigRz}. As the derivatives are obtained based on only 4 radial bins, they have large error bars. See \cite{hagen2019} for more detailed trends in the tilt angle of the velocity ellipsoids in a larger spatial range. Except for several bins in the south, which have few stars, the derivatives of $\sigma^{2}_{Rz}$ can be roughly fitted with a linear function. The zero point of the derivatives is at $z \sim 0.5$ kpc, rather than z = 0 kpc. The difference in the derivative of $\sigma^{2}_{Rz}$ will then result in different contributions to the $\sigma_{z}$ profiles for the northern and southern subsamples.

%%%%%%%%%%% Fig of tilt term
%%%%%%%%%%%

%%%%%%%%%%%%%%
\section{Results}
\label{sec_result}
%%%%%%%%%%%%%%
\subsection{Results under a Gaussian prior on $\Sigma_{\star}$}
\label{ssec_prior}
With the vertical Jeans equation and mass models described in Section \ref{sec_method}, we use MCMC to obtain estimates of the four parameters. For $z_{\rm h}$, $\rho_{\rm dm}$ and $\sigma_{z}(z_{0})$, we apply almost non-informative priors by constraining them in ranges of $0 < z_{\rm h} < 1000$ pc, $0 < \rho_{\rm dm} < 0.05\ {\rm M_{\odot}\,pc}^{-3}$ and $10 < \sigma_{z}(z_{0}) < 30$ km s$^{-1}$. However, for $\Sigma_{\star}$, we use a Gaussian prior with a mean of 37.0 ${\rm M}_{\odot}\,{\rm pc}^{-2}$ and a dispersion of 5.3 ${\rm M}_{\odot}\,{\rm pc}^{-2}$. This prior is derived from a compilation of several previous works, which are based on different methods including stellar census \citep{flynn2006}, Jeans modelling \citep[e.g.][]{kuijken1989b, zhang2013} and action-based distribution function \citep{bovy2013} etc. These works are assumed to be independent measurements extracted from a true value with an intrinsic dispersion. We apply a Hierarchical Bayesian Model to these measurements to derive the Gaussian prior used in this work. Measurements together with references and the derived prior are shown in Table \ref{tab_prior}.

%%%%%%%%%%%%%%
\begin{table*}
  \centering
  \caption{Compilation of measurements of the total stellar surface density $\Sigma_{\star}$ (second column) and of the stellar volume density $\rho_{\star,0}$ on the Galctic plane (fourth column). The values and their errors derived from the Hierarchical Bayesian analyses of the listed measurements are shown in the bottom row.}
  \label{tab_prior}
  \begin{threeparttable}
	  \begin{tabular}{l c l c}
	  \hline
	  Reference & $\Sigma_{\star}$ [${\rm M}_{\odot}\,{\rm pc}^{-2}$] & Reference & $\rho_{\star,0}$ [${\rm M}_{\odot}\,{\rm pc}^{-3}$] \\
	  \hline
	  \citet{kuijken1989b} & $35.0 \pm 5.0$ & \citet{holmberg2000} & $0.044 \pm 0.0044$\tnote{4}  \\
	  \citet{flynn2006} & $35.5 \pm 3.6$\tnote{1} & \citet{chabrier2001} & $0.045 \pm 0.003$ \\
	  \citet{bovy2013} & $38.0 \pm 4.0$ & \citet{flynn2006} & $0.042 \pm 0.0042$\tnote{1} \\
	  \citet{zhang2013} & $43.6 \pm 5.0$\tnote{2} & \citet{bovy2017} & $0.0472 \pm 0.003$\tnote{5} \\
	  \citet{read2014} & $37.2 \pm 1.2$\tnote{3} & \citet{schutz2018} & $0.043 \pm 0.004$\tnote{6}  \\
	  \citet{sivertsson2018} & $33.2 \pm 5.3$ & \citet{xiang2018} & $0.0536 \pm 0.0007$  \\
	  \citet{xiang2018} & $36.8 \pm 0.5$ & \multicolumn{1}{c}{--} & --  \\
	  This compilation & $37.0 \pm 5.3$ & This compilation & $0.0468 \pm 0.0050$ \\
	  \hline
	  \end{tabular}
	  \begin{tablenotes}
	    \footnotesize
	    \item[1] In the stellar census, uncertainties on the densities of all the stellar components are $\sim$ 10\%.
	    \item[2] A 300 pc scale height is used to extrapolate their measurement $42\pm 5$ ${\rm M}_{\odot}\,{\rm pc}^{-2}$ at 1.0 kpc to the total value here.
	    \item[3] This review article compiles several literature results.
	    \item[4] A typical 10\% uncertainty in luminosity is assumed.
	    \item[5] Their result 0.040 ${\rm M}_{\odot}\,{\rm pc}^{-3}$ is just for main sequence stars. An amount of 0.0072 ${\rm M}_{\odot}\,{\rm pc}^{-3}$ is taken from \cite{mckee2015} for brown and white dwarfs.
	    \item[6] This article compiles results from \cite{flynn2006}, \cite{mckee2015} and \cite{read2014}.
	  \end{tablenotes}
  \end{threeparttable}
\end{table*}
%%%%%%%%%%%%%%

\begin{figure*}
   \centering
   \includegraphics[width=0.8\textwidth]{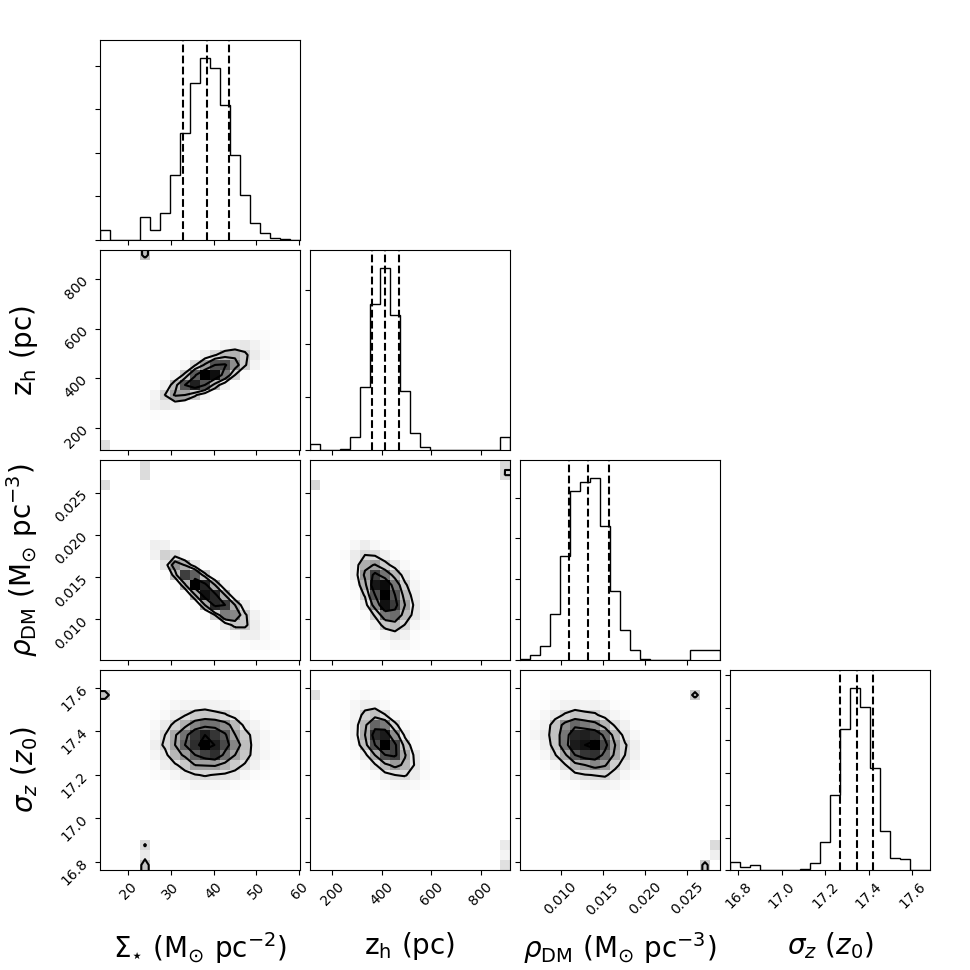}
   \caption{The probability density functions (PDFs) of the four model parameters from MCMC. The parameters from the left to the right are $\Sigma_{\star}$, $z_{\rm h}$, $\rho_{\rm dm}$ and $\sigma_{z}(z_{0})$, respectively. The parameters from the top to the bottom are $z_{\rm h}$, $\rho_{\rm dm}$ and $\sigma_{z}(z_{0})$, respectively. The solid and dashed lines in each histogram indicate the median, 16th and 84th percentiles of the 1D marginalized PDF of each parameter.}
   \label{fig_tri}
\end{figure*}

The posterior probability density functions (PDFs) of the four parameters under the Gaussian prior of $\Sigma_{\star}$ are shown in Fig. \ref{fig_tri}. The estimates of the parameters and errors, listed in the second column of Tabel \ref{tab_para}, are taken from the median, 16th and 84th percentiles of each 1D marginalized PDF over other three parameters. The model predicted velocity dispersion is over-plotted as the red solid line in Fig. \ref{fig_sigmav}. In the MCMC, after the initial iterations, each set of parameters can give a model velocity dispersion profile. The red line is then the median profile of all the model profiles. The 1$\sigma$ region, indicated by the red dashed lines in Fig. \ref{fig_sigmav}, is calculated from the 16th and 84th percentiles of these profiles. Our result predicts a $\sigma_{z}$ profile consistent with the observed profile, as shown in Fig. \ref{fig_sigmav}.

%%%%%%% figure of PDF of priors
%%%%%%%

%%%%%%%
\begin{table*}
  \centering
  \caption{Parameters from MCMC under different priors. Values under Gaussian priors on the total stellar surface density ($P(\Sigma_{\star}) \sim {\rm N}(37.0,5.3^{2})\ {\rm M}_{\odot}\,{\rm pc}^{-2}$), on the stellar volume density on the Galactic plane ($P(\rho_{\star,0}) \sim {\rm N}(0.0468,0.0050^{2})\ {\rm M}_{\odot}\,{\rm pc}^{-3}$) and a non-informative prior are listed in the second, third and fourth columns, respectively. The parameter $\rho_{\star,0}$ is also given in the bottom row.}
  \label{tab_para}
  \begin{tabular}{c c c c}
     \hline
     Parameter  &  $P(\Sigma_{\star}) \sim {\rm N}(37.0,5.3^{2})\ {\rm M}_{\odot}\,{\rm pc}^{-2}$  &  $P(\rho_{\star,0}) \sim {\rm N}(0.0468,0.0050^{2})\ {\rm M}_{\odot}\,{\rm pc}^{-3}$  &  Non-informative prior  \\
     \hline
     \vspace{1.2mm}
     $\Sigma_{\star}$ [${\rm M}_{\odot}\,{\rm pc}^{-2}$] & $38.4_{-5.4}^{+5.2}$ & $57.1_{-18.1}^{+15.4}$ & $62.6_{-18.9}^{+13.2}$ \\
     \vspace{1.2mm}
     $z_{\rm h}$ [pc] & $413_{-53}^{+57}$ & $572_{-157}^{+124}$ & $591_{-141}^{+114}$ \\
     \vspace{1.2mm}
     $\rho_{\rm dm}$ [${\rm M}_{\odot}\,{\rm pc}^{-3}$] & $0.0133_{-0.0022}^{+0.0024}$ & $0.0071_{-0.0043}^{+0.0059}$ & $0.0049_{-0.0037}^{+0.0061}$ \\
     \vspace{1.2mm}
     $\sigma_{z}(z_{0})$ [${\rm km}\,{\rm s}^{-1}$] & $17.3_{-0.1}^{+0.1}$ & $17.3_{-0.1}^{+0.1}$ & $17.3_{-0.1}^{+0.1}$ \\
     \vspace{1.2mm}
     $\rho_{\star,0}$ [${\rm M}_{\odot}\,{\rm pc}^{-3}$] & $0.0468_{-0.0040}^{+0.0039}$ & $0.0499_{-0.0037}^{+0.0035}$ & $0.0521_{-0.0050}^{+0.0039}$ \\
     \hline
  \end{tabular}
\end{table*}
%%%%%%%%

The model predicts $\sigma_{z}(z_{0}) = 17.3 \pm 0.1$ km s$^{-1}$, which is well constrained with a quite small error. This value is a little different from the binned value $17.7 \pm 0.2$ km s$^{-1}$. The small uncertainty in $\sigma_{z}(z_{0})$ is due to its exponentially increasing contribution to the model velocity dispersion, shown as the second term in the right-hand side of Eq. \ref{eq_sigz}. Besides, $\sigma_{z}(z_{0})$ almost has no correlation with other three parameters, except for a small anti-correlation with $z_{\rm h}$. Higher $\sigma_{z}(z_{0})$ will result in a steeper exponential increase in the model velocity dispersion profile. Thus a lower scale height is needed to compensate the velocity dispersion of low-$z$ region in order to get an overall gradually increasing profile.

\begin{figure}
   \centering
   \includegraphics[width=0.9\columnwidth]{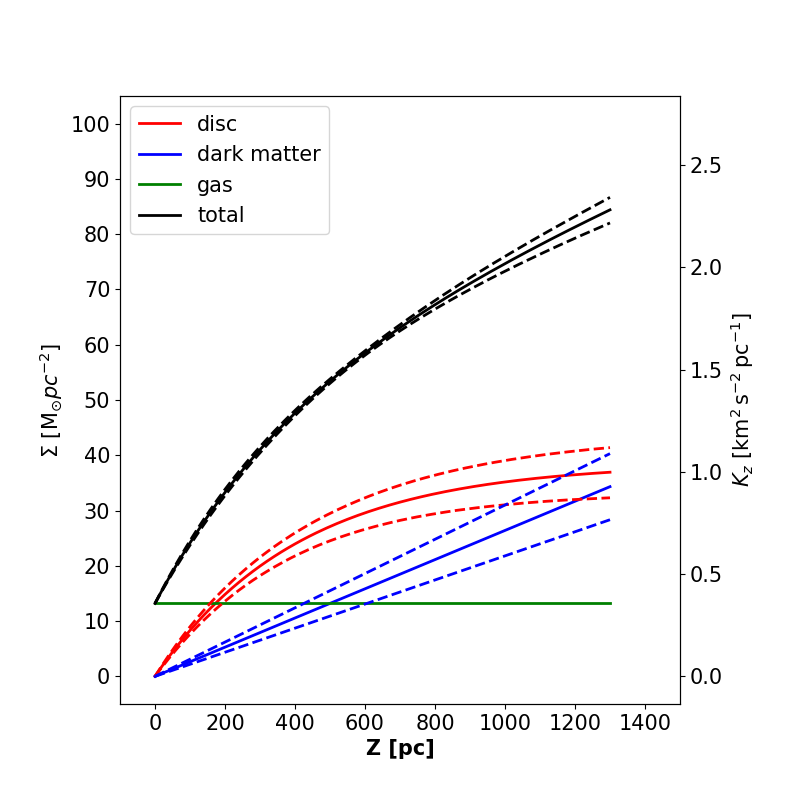}
   \caption{The surface density profile, or equally the $K_{z}$ force predicted by MCMC. The black solid line is the total surface density profile. The coloured solid lines show the surface density profiles of the stellar disc (red), the razor thin gas disc (green) and the dark matter (blue), respectively. The dashed lines show the corresponding 1$\sigma$ regions.}
   \label{fig_kz}
\end{figure}

As shown in Fig. \ref{fig_tri}, $\Sigma_{\star}$ has a strong anti-correlation with $\rho_{\rm dm}$, and a strong positive correlation with $z_{\rm h}$. These correlations can be explained from the surface density profiles shown in Fig. \ref{fig_kz}. As $\rho_{\rm dm}$ increases, both $\Sigma_{\star}$ and its growth rate decrease due to the constraint from the total surface density. This means the disc will become flat more quickly, and thus the scale height $z_{\rm h}$ will be smaller. Although $z_{\rm h}$ and $\rho_{\rm dm}$ seem to be positively correlated from Eq. \ref{eq_kz_int}, their correlation is weak and is influenced by $\Sigma_{\star}$.

The model predicted $K_{z}$ force, or equally the total surface density profile, is shown as the black solid line in Fig. \ref{fig_kz}. The contributions from different components are shown as coloured solid lines. The 1$\sigma$ errors are calculated similarly to that of the model velocity dispersion, and are shown in Fig. \ref{fig_kz} as dashed lines for the total mass, stellar disc and dark matter. The total surface density is well constrained, with $\Sigma_{{\rm tot},\ |z|<1.0{\rm kpc}} = 74.7_{-1.4}^{+1.4}\ {\rm M}_{\odot}\,{\rm pc}^{-2}$ for the value up to 1 kpc. However, the total stellar surface density has larger errors with a value of $\Sigma_{\star,\ |z|<1.0{\rm kpc}} = 35.0_{-4.4}^{+4.0}\ {\rm M}_{\odot}\,{\rm pc}^{-2}$. The relatively larger errors are due to the degeneracy between the stellar disc and the dark matter, the latter having a volume density of $\rho_{\rm dm} = 0.0133_{-0.0022}^{+0.0024}\ {\rm M}_{\odot}\,{\rm pc}^{-3}$. The contribution from the razor thin gas disc is fixed as a constant $\Sigma_{\rm gas} = 13.2\ {\rm M}_{\odot}\,{\rm pc}^{-2}$.

%%%%%%%%%% fig of Sigma_z or Kz
%%%%%%%%%%%

\subsection{Results of other priors}
\label{ssec_otherp}
In addition to the Gaussian prior on $\Sigma_{\star}$, we also try to use a Gaussian prior on the stellar volume density on the Galactic plane ($\rho_{\star,0}$) or non-informative priors on all parameters. The results based on these two kinds of priors are listed in Table \ref{tab_para}. The prior on $\rho_{\star,0}$ is $P(\rho_{\star,0}) \sim {\rm N}(0.0468,0.0050^{2})\ {\rm M}_{\odot}\,{\rm pc}^{-3}$, which is also from a compilation of several previous works listed in Table \ref{tab_prior}. The reason for using this prior is that we regard works based on stellar census \citep[e.g.][]{flynn2006, bovy2017, schutz2018} as able to give quite reliable estimates of the volume densities of stellar components in the solar vicinity. The Gaussian prior on the local stellar volume density is usually used in works which apply a distribution function to construct the potential \citep[e.g.][]{buch2019, widmark2019}.

The dark matter density measured using the prior $P(\rho_{\star,0}) \sim {\rm N}(0.0468,0.0050^{2})\ {\rm M}_{\odot}\,{\rm pc}^{-3}$ is $0.0071_{-0.0043}^{+0.0059}\ {\rm M}_{\odot}\,{\rm pc}^{-3}$, which is smaller than the result of the Gaussian prior on $\Sigma_{\star}$. Consequently, $\Sigma_{\star}$ and $z_{\rm h}$ are larger. Nevertheless, due to the large error bars in the results of the Gaussian prior on $\rho_{\star,0}$, these measurements are consistent within 1$\sigma$. This prior is weaker than the Gaussian prior on $\Sigma_{\star}$.

The non-informative priors on all parameters lead to quite a small $\rho_{\rm dm}$ and high $\Sigma_{\star}$ and $z_{\rm h}$. This result is similar to the mock data 3 result in Section 4.3 of \cite{xia2016}, which has a data range of $0 < z < 1000$ pc. The most prominent feature is that the local dark matter density has a peak close to zero, as shown in Fig. 11 of \cite{xia2016}. The reason is that the data lack stars with high $z$, where the contribution of the dark matter to the mass profile becomes significant. Thus, in the likelihood calculation of MCMC, the resulting model parameter PDF is biased to show a low $\rho_{\rm dm}$. Though $\rho_{\rm dm}$ is biased to a lower value and $\Sigma_{\star}$ and $z_{\rm h}$ have quite large uncertainties, the local stellar volume density $\rho_{\star,0}$ is well constrained with an error about 0.005 ${\rm M}_{\odot}\,{\rm pc}^{-3}$, comparable to previous works listed in Table \ref{tab_prior}. The positive correlation between $\Sigma_{\star}$ and $z_{\rm h}$ could be mainly due to $\rho_{\star,0}$. Thus, $\rho_{\star,0}$ is better constrained than $\Sigma_{\star}$ and $z_{\rm h}$.

\section{Discussions}
\label{sec_discuss}
%%%%%%%%%%%%%%%
\subsection{Comparisons with previous works}
\label{ssec_comp}
The local dark matter density $\rho_{\rm dm}$ given in this work is $0.0133_{-0.0022}^{+0.0024}\ {\rm M}_{\odot}\,{\rm pc}^{-3}$. This value is consistent with several previous works, such as \cite{bienayme2014}, \cite{piffl2014}, \cite{budenbender2015}, \cite{sivertsson2018}, but is inconsistent with works such as \cite{bovy2013}, \cite{zhang2013}, \cite{hagen2018}. These inconsistencies arise from complicated reasons including the sample used, the data reduction, the methods applied, and the simplifications or priors utilized.

The most closely related works, \cite{zhang2013} and \cite{xia2016} gave quite different measurements for the local dark matter density, $0.0065 \pm 0.0023$ and $0.018 \pm 0.0054$ ${\rm M}_{\odot}\,{\rm pc}^{-3}$, respectively. They both adopted the non-informative prior. In addition, \cite{zhang2013} combined together metal-rich, intermediate-metallicity, and metal-poor subsamples in their model. This work obtains an intermediate value with a smaller uncertainty. Though our sample size is much larger than the two samples used by \cite{zhang2013} and \cite{xia2016}, the smaller errors are mainly due to the Gaussian prior on $\Sigma_{\star}$ applied here. However, differences in the median values are not only due to the prior, but also to the different data ranges. We will discuss this later in more detail in Section \ref{ssec_error}.

The estimated density scale height $z_{\rm h}$ here is $413_{-53}^{+57}$ pc, which is slightly larger than the usually quoted value of 300 pc for the thin disc but is consistent with the estimated disc thickness of 500 pc within the error bars \citep{binney2008}. The stellar local volume density $\rho_{\star,0}$ can be derived from $\rho_{\star,0} = \frac{\Sigma_{\star}}{2z_{\rm h}}$. This work obtains $\rho_{\star,0} = 0.0468_{-0.0040}^{+0.0039}\ {\rm M}_{\odot}\,{\rm pc}^{-3}$, which is consistent with previous works listed in Table \ref{tab_prior}. Due to the strong positive correlation between $\Sigma_{\star}$ and $z_{\rm h}$, works applying $z_{\rm h} = 300$ pc or fixing the scale heights of thin and thick discs actually took a stronger prior than this work \citep[e.g.][]{bienayme2014, budenbender2015}. As a consequence, they usually have smaller uncertainties.

%%%%%%%%%%%%%%%
\subsection{$\rho_{\rm dm}$ vs. $\phi$}
\label{ssec_rhophi}
\begin{figure}
   \centering
   \includegraphics[width=0.9\columnwidth]{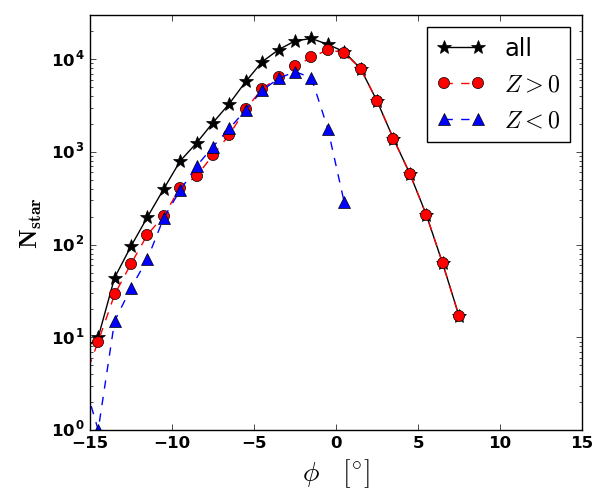}
   \caption{Stellar number distribution as a function of azimuthal angle. The sample accounts for all the selection criteria listed in Section \ref{ssec_dsc} except for the constraint on the azimuthal angle. The black stars show the distribution of the total sample, while the red dots and the blue triangles show the distributions of stars with $z > 0$ and $z < 0$, respectively.}
   \label{fig_phi_dis}
\end{figure}

As our sample covers a large azimuthal angle range, we separate the sample into subsamples with different $\phi$ and $z$ ranges. We also explore a larger azimuthal angle range according to the data coverage as shown in Fig. \ref{fig_phi_dis}. All the selection criteria, except for the azimuthal angle, are the same as the criteria listed in Section \ref{ssec_dsc}. Note that the Sun is placed at ($-$8.34, 0., 0.027) kpc in the Galactic Cartesian coordinates system. Thus $\phi < 0$ is the direction of the Galactic rotation. There are few stars in the southern sky with $\phi >0$ and in the region with $\phi > 5^{\circ}$, as shown in Fig. \ref{fig_phi_dis}. Thus, we separate all the stars into eight subsamples: 
\\

(1) $-5^{\circ} < \phi < 5^{\circ}$, i.e. the sample used in previous analyses;

(2) $-5^{\circ} < \phi < 5^{\circ}$ and $z > 0$;

(3) $-5^{\circ} < \phi < 5^{\circ}$ and $z < 0$;

(4) $0^{\circ} < \phi < 5^{\circ}$, where the few stars with $z < 0$ are excluded;

(5) $-5^{\circ} < \phi < 0^{\circ}$;

(6) $-5^{\circ} < \phi < 0^{\circ}$ and $z > 0$;

(7) $-5^{\circ} < \phi < 0^{\circ}$ and $z < 0$;

(8) $-10^{\circ} < \phi < -5^{\circ}$, where the few stars with $z < 100$ pc are excluded.

%%%%%%%%%%% Fig of phi distribution
%%%%%%%%%%%

\begin{figure}
   \centering
   \includegraphics[width=0.9\columnwidth]{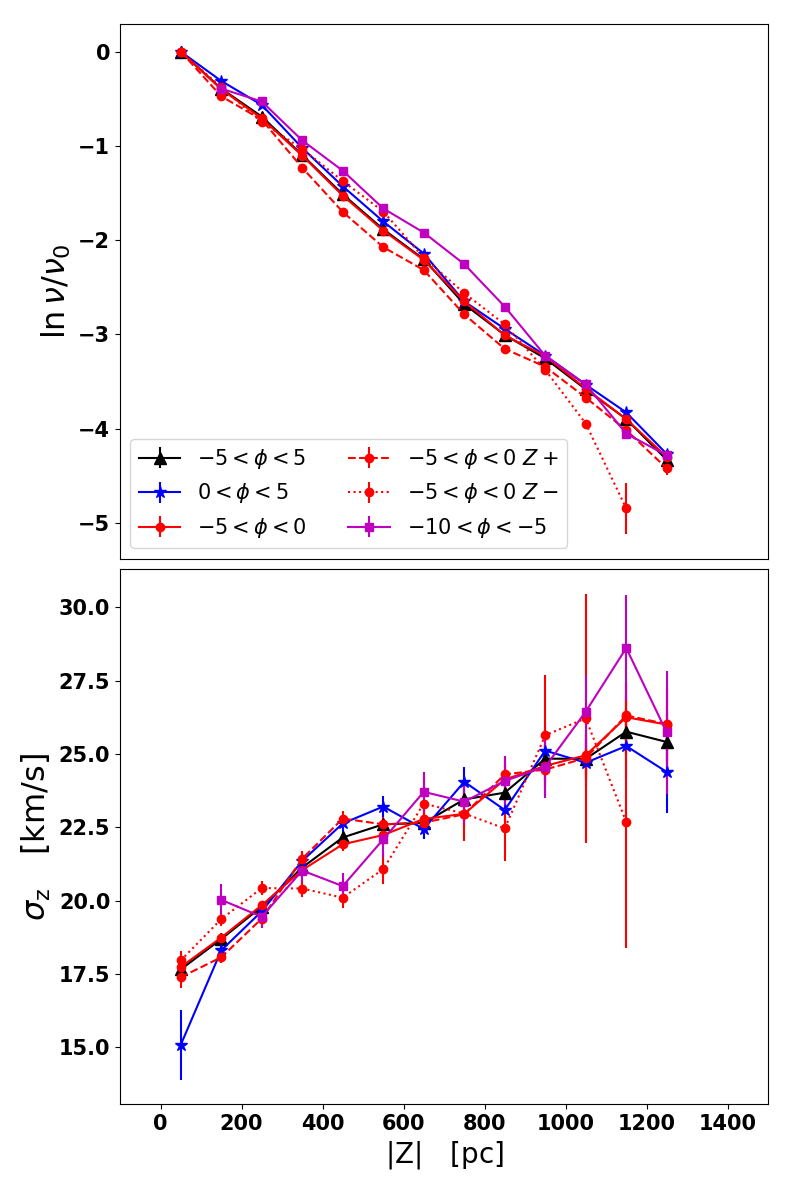}
   \caption{The number density profiles (upper panel) and the vertical velocity dispersion profiles (lower panel) of the subsamples. The number density distributions are selection effects corrected and normalized to the first bins. The black triangles, the blue stars, the red dots and the magenta squares, connected with solid lines, show the profiles of stars with $-5^{\circ} < \phi < 5^{\circ}$, $0^{\circ} < \phi < 5^{\circ}$, $-5^{\circ} < \phi < 0^{\circ}$ and $-10^{\circ} < \phi < -5^{\circ}$, respectively. The red dashed line shows the profile of stars with $-5^{\circ} < \phi < 0^{\circ}$ and $z > 0$, while the red dotted line displays the profile of stars with $-5^{\circ} < \phi < 0^{\circ}$ and $z < 0$.}
   \label{fig_phi_nusig}
\end{figure}

The number density profiles and velocity dispersion profiles of the eight subsamples are shown in the upper and lower panels of Fig. \ref{fig_phi_nusig}. The logarithmic number density profiles can be linearly well fitted and have similar scale heights except for the last two bins in subsample (7). These subsamples also show similar velocity dispersion profiles except for the noise in the high $z$ region as a result of the low star numbers. Subsample (8) suffers from a small sample size, and shows a fluctuating $\sigma_{z}$ profile. An obvious discrepancy happens between subsamples (6) and (7). Subsample (6) (i.e. the stars in the northern sky) shows a plateau in the region of $400 < z < 600$ pc, while subsample (7) (the stars in the southern sky) shows a plateau in the region of $200 < z < 400$ pc. These plateaux will significantly influence the determination of the disc scale height $z_{\rm h}$ and thus the $\rho_{\rm dm}$.

%%%%%%%%%%% Fig of nuz & sigz of phis
%%%%%%%%%%%

%%%%%%%%%%% Fig of paras of phis
\begin{figure}
   \centering
   \includegraphics[width=0.96\columnwidth]{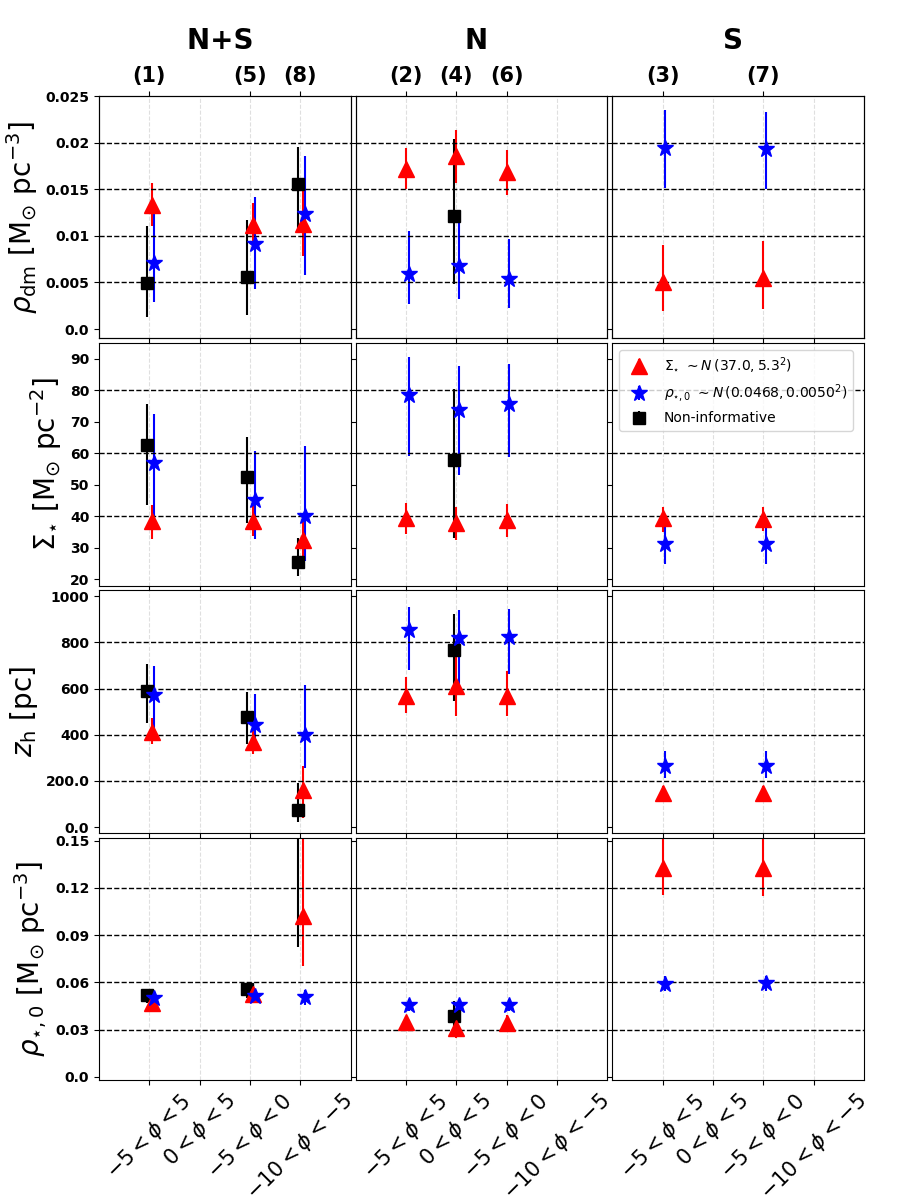}
   \caption{Model predicted parameters for different priors and subsamples. The parameters from the top to the bottom are $\rho_{\rm dm},\ \Sigma_{\star},\ z_{\rm h},\ \rho_{\star,0}$, respectively. The horizontal coordinates are four azimuthal angle ranges. The eight subsamples are separated into total (left column), northern (middle column) and southern (right column) subsamples. The indices of the subsamples, the same as those in Section \ref{ssec_rhophi}, are also labelled at the top of the panels in the top row. The three different priors, i.e. a Gaussian prior on $\Sigma_{\star}$, a Gaussian prior on $\rho_{\star,0}$ and a non-informative prior, are shown as the red triangles, the blue stars and the black squares, respectively.}
   \label{fig_rhodm_phi}
\end{figure}
%%%%%%%%%%%

The model parameters given by the eight subsamples are shown in Fig. \ref{fig_rhodm_phi}. Subsamples (4) and (6), both in the northern sky, result in similar values of $\rho_{\rm dm}$. The model predicted $\rho_{\rm dm}$, under the Gaussian prior on $\Sigma_{\star}$, are $0.0185_{-0.0028}^{+0.0028}\ {\rm M}_{\odot}\,{\rm pc}^{-3}$ and $0.0168_{-0.0024}^{+0.0024}\ {\rm M}_{\odot}\,{\rm pc}^{-3}$ for subsamples (4) and (6), respectively. These values are consistent with the value $0.018 \pm 0.0054\ {\rm M}_{\odot}\,{\rm pc}^{-3}$ from \cite{xia2016}, which also uses stars in the northern sky. The velocity dispersion profile of \cite{xia2016} shows a clear dip at $z = 650$ pc, which is similar to our subsample (4) and may be related to the plateau in subsample (6). However, $\rho_{\rm dm}$ derived from the Gaussian prior on $\rho_{\star,0}$ are much smaller than that from the Gaussian prior on $\Sigma_{\star}$. The predicted values of $\rho_{\rm dm}$ are $0.0067_{-0.0035}^{+0.0050}\ {\rm M}_{\odot}\,{\rm pc}^{-3}$ and $0.0054_{-0.0031}^{+0.0042}\ {\rm M}_{\odot}\,{\rm pc}^{-3}$ for subsamples (4) and (6), respectively. These lower $\rho_{\rm dm}$ are not due to the samples used, but due to the prior applied. The Gaussian prior on $\Sigma_{\star}$ seems to give a lower $\rho_{\star,0}$ and thus a higher $\rho_{\rm dm}$, as shown in Fig. \ref{fig_rhodm_phi}.

Subsamples (1), (5) and (8) contain stars both in the northern and southern sky. They obtain quite consistent measurements on $\rho_{\rm dm}$ for both priors, though the subsample (8) suffers from a small sample size and large noises in the $\nu_{z}$ and $\sigma_{z}$ profiles. The consistency in the results from subsamples (1), (5) and (8), or from the subsamples (4) and (6) indicates that the dark matter densities are quite similar in an azimuthal angle range of $-10^{\circ} < \phi < 5^{\circ}$.

\subsection{Results from the northern and southern sky}
\label{ssec_asy_NS}
The results obtained from different azimuthal angle ranges are roughly similar. However, the measured $\rho_{\rm dm}$ and $z_{\rm h}$ show a large discrepancy between the northern subsample with $z > 0$ and the southern subsample with $z < 0$, as shown by the middle and right columns in Fig. \ref{fig_rhodm_phi}. In this section, we take subsamples (1), (2) and (3) for a more detailed discussion of the north and south asymmetry.

\subsubsection{Profiles in the north and south}
\label{sssec_NS_vdis}
\begin{figure*}
   \centering
   \includegraphics[width=0.9\textwidth]{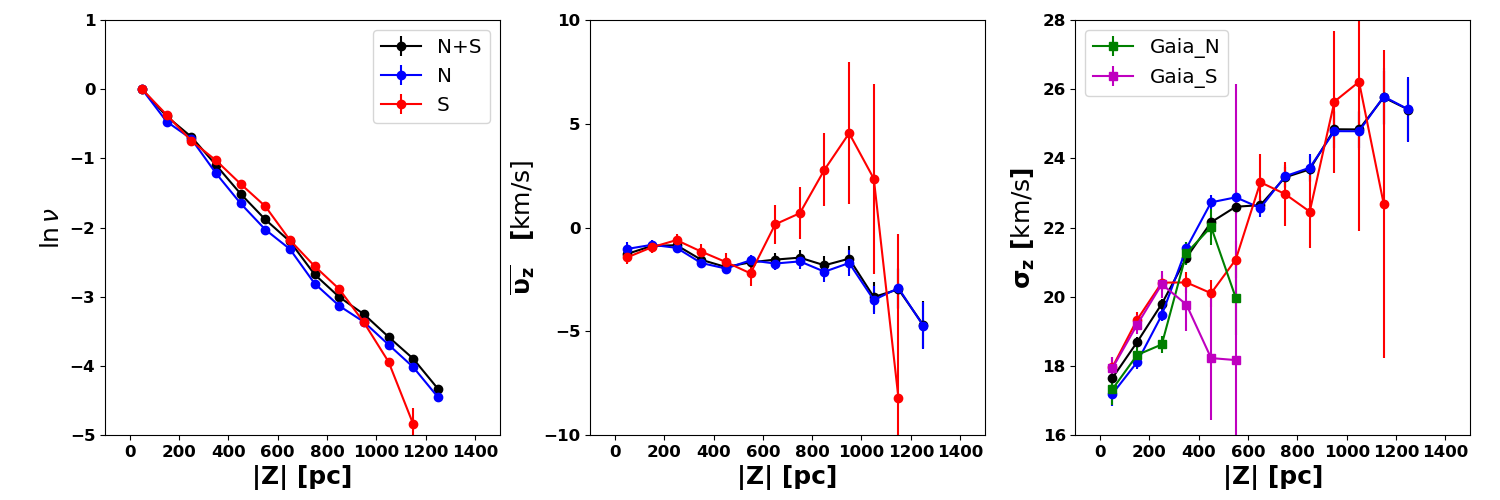}
   \caption{The number density profiles (left panel), mean vertical velocity profiles (middle panel) and vertical velocity dispersion profiles (right panel) of our subsamples. The black, blue and red dots stand for the total subsample (i.e. the subsamples `(1)'), the northern subsample (subsample `(2)') and the southern subsample (subsample `(3)'), respectively. In the right panel, the northern and southern $\sigma_{z}$ profiles derived from Gaia radial velocities are shown as green and magenta squares respectively.}
   \label{fig_NS_profiles}
\end{figure*}

The number density profiles, mean vertical velocity profiles and vertical velocity dispersion profiles of subsamples (1) (the total subsample), (2) (the northern subsample) and (3) (the southern subsample) are shown in Fig. \ref{fig_NS_profiles}. The number density profiles are quite similar, while the velocity dispersion and mean velocity profiles show some differences. There are two plateaux in the northern and southern $\sigma_{z}$ profiles at different vertical height, which result in different estimates for parameters.

For the results obtained using the Gaussian prior of $\Sigma_{\star}$, shown as the red triangles in Fig. \ref{fig_rhodm_phi}, the northern subsample has larger $\rho_{\rm dm}$ and $z_{\rm h}$ than the southern subsample. The former yields a $\rho_{\rm dm}$ of $0.0172_{-0.0022}^{+0.0022}\ {\rm M}_{\odot}\,{\rm pc}^{-3}$, while the latter has a value of $0.005_{-0.0031}^{+0.0039}\ {\rm M}_{\odot}\,{\rm pc}^{-3}$. This difference comes from the difference in the velocity dispersion profiles. The position of the plateau in $\sigma_{z}$ profile gives a strong constraint on the disc scale height, because the contribution from the stellar disc to the $K_{z}$ force is approximately flat over about two scale heights. Thus, the northern subsample, which shows a plateau in the region of $400 < z < 700$ pc, has $z_{\rm h} = 566_{-73}^{+85}\ {\rm pc}$. The southern subsample has $z_{\rm h} = 148_{-27}^{+30}\ {\rm pc}$ due to the plateau in the region of $200 < z < 500$ pc. Consequently, under the same Gaussian prior on $\Sigma_{\star}$, the southern subsample has a larger $\rho_{\star,0}$ and a smaller $\rho_{\rm dm}$ than the northern subsample. For the results derived from the Gaussian prior on $\rho_{\star,0}$, the larger scale height results in a larger $\Sigma_{\star}$ and thus a smaller $\rho_{\rm dm}$ for the northern subsample, shown as the blue stars in Fig. \ref{fig_rhodm_phi}. The combined subsample (1) has a smoother velocity dispersion profile and a local dark matter density ($0.0133_{-0.0022}^{+0.0024}\ {\rm M}_{\odot}\,{\rm pc}^{-3}$) positioned between the values of the two separated subsamples.

%%%%%%%%%%% Fig of nuz, sigz, mv
%%%%%%%%%%%

%%%%%%%%%%% Fig of vz distribution
\begin{figure*}
   \centering
   \includegraphics[width=0.9\textwidth]{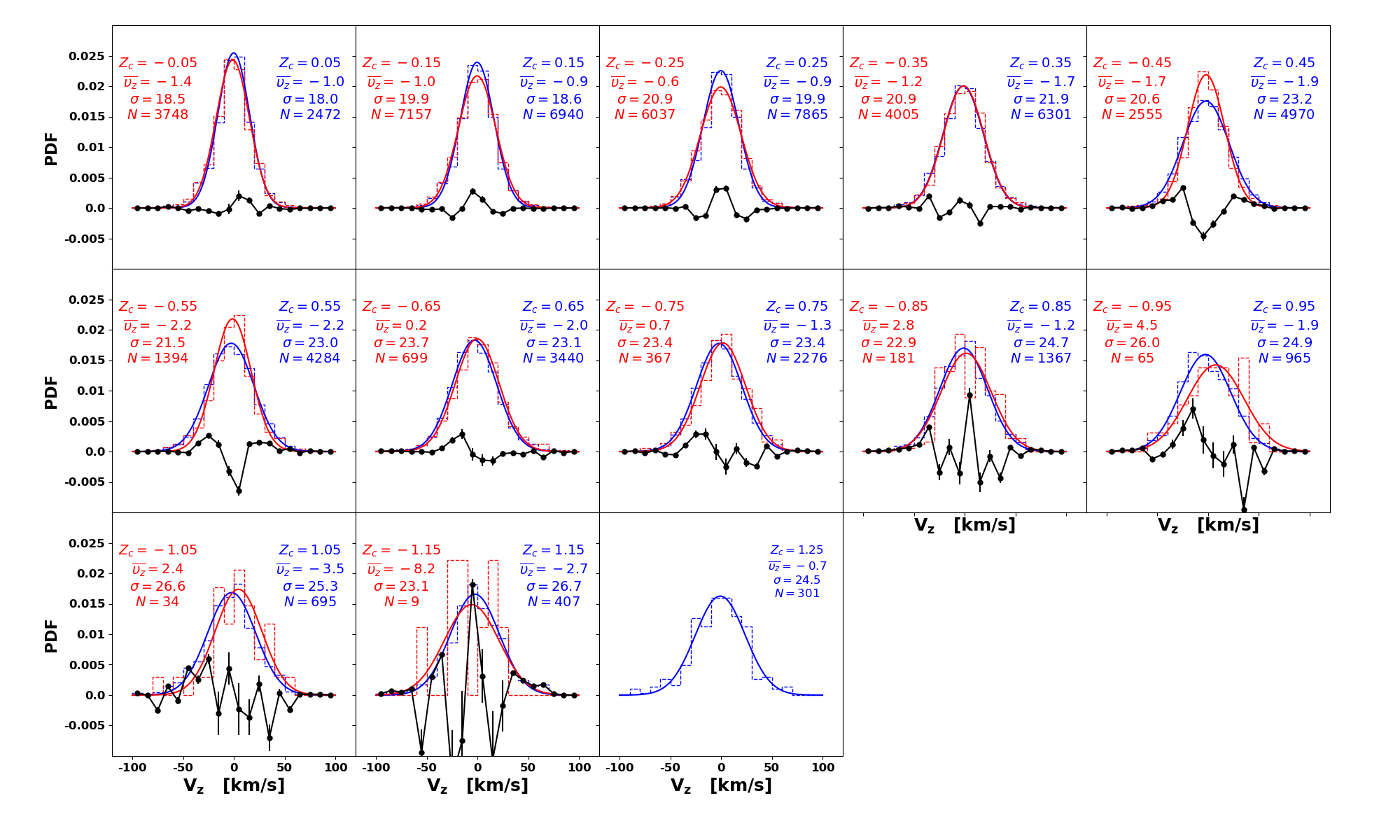}
   \caption{Velocity distributions of the northern (blue lines) and southern (red lines) subsamples. The velocities are derived from the LAMOST radial velocity measurements. The solid lines are the Gaussian fits of the dashed histograms. The bin centre, mean velocity, velocity dispersion are labelled in units of kpc, ${\rm km\,s^{-1}}$, and ${\rm km\,s^{-1}}$, respectively. The black lines show the differences between the northern and southern subsamples, with errors from numerical noise.}
   \label{fig_NS_vdis}
\end{figure*}
%%%%%%%%%%%

The velocity distributions of the northern and southern subsamples in each bin are shown in Fig. \ref{fig_NS_vdis}. The bin width is 100 pc. The stellar counts in each bin are also labelled in the figure. The lower six bins all contain several thousands of stars, which ensure the accuracy of the histograms. The histograms can be well fitted with a single Gaussian function. The velocity dispersion of the southern subsample is a little larger than that of the northern subsample for $z < 300$ pc. This situation is then reversed for $300 < z < 600$ pc. The difference reaches a significant value of 2.6 ${\rm km\,s^{-1}}$ at $z = 450$ pc. In the outer region, the velocity dispersions are consistent given the large errors of the southern subsample. Due to the large sample size and the well fitted Gaussian distribution, we regard the difference between the northern and southern subsamples as reliable and significant, and the contamination of the thick component stars as trivial.

The velocity dispersion profiles derived from about 26,000 stars, which have Gaia radial velocity measurements, are also plotted in Fig. \ref{fig_NS_vdis}. The profiles are similar to those derived from LAMOST radial velocity measurements. The difference at $z = 450$ pc is also significant. In addition, in our preliminary analyses of giant stars and samples with $5.0 < M_{G} < 6.0$, similar differences between the northern and southern $\sigma_{z}$ profiles are found.

Similar structures in the southern sky are also found in several previous works. \cite{bienayme2014} used $\sim 4600$ red clump stars in the southern sky to measure the local dark matter density. Their $\sigma_{z}$ profile shows a dip at $z \sim 400$ pc, as shown in their Fig. 8. This dip could be related to the plateau of our southern subsamples. In addition, their $K_{z}$ force profile, derived directly from the $\sigma_{z}$ profile, shows a rapid increase for $z < 400$ pc. This rapid increase will result in quite a small scale height ($z_{\rm h}$) if it modelled as a free parameter. \cite{garbari2012} re-examined $\sim 2000$ K dwarf stars in the southern sky, taken from \cite{kuijken1989b}. There is an obvious plateau at $400 < z < 700$ pc in their $\sigma_{z}$ profile, as shown in their Fig. 5. However, their distances are obtained from a relationship between the metallicity, the vertical distance z and the V-band absolute magnitude. Thus, their $\sigma_{z}$ profile could be systematically shifted, and the plateau in their $\sigma_{z}$ profile could be related to that of our southern $\sigma_{z}$ profile. \cite{hagen2018} investigated the kinematics of red clump stars by combining data from TGAS and RAVE. Their $\sigma_{z}$ profiles of thin disc samples show plateaux at $300 < z < 500$ pc, as shown in their Figs. 5 and 6. Their samples contain both northern and southern stars. Nevertheless, they have more stars in the southern sky according to their Fig. 2 showing stellar spatial distribution. Thus, their plateau in $\sigma_{z}$ is similar to ours.

The mean velocity profiles of the northern and southern subsamples also show some differences. However, that difference is different from the one in the velocity dispersion profiles. At $z < 600$ pc, the mean velocity profiles are similar, and show a bulk motion of about -1.5 ${\rm km\,s^{-1}}$. In the higher region, the tracer population shows a motion consistent with disc compression. The mean velocity has a difference of $\sim$ 7 ${\rm km\,s^{-1}}$ at $z = 950$ pc. The difference in the mean velocity profiles could be another sign of dis-equilibrium of the local disc.

\subsubsection{Results with the tilt term}
\label{sssec_NS_tilt}
%%%%%%%
\begin{table*}
  \centering
  \caption{Parameters derived from the subsamples with the tilt term taken into consideration and with the errors of bins enlarged. In both situations, the Gaussian prior $P(\Sigma_{\star}) \sim {\rm N}(37.0,5.3^{2})\ {\rm M}_{\odot}\,{\rm pc}^{-2}$ is utilised. Results for the total, the northern and the southern subsamples are listed in the third, fourth and fifth columns, respectively.}
  \label{tab_tilt_Lerr}
  \begin{tabular}{p{2.cm}| c c c c}
     \hline
      & Parameter  &  Total  &  North  &  South  \\
     \hline
     \vspace{1.2mm}
     \multirow{3}{*}{ with Tilt Term } & $\Sigma_{\star}$ [${\rm M}_{\odot}\,{\rm pc}^{-2}$] & \multirow{3}{*}{ -- } & $38.4_{-5.1}^{+5.1}$ & $38.6_{-4.3}^{+4.0}$ \\
     %\cline{2-2}
     \vspace{1.2mm}
     & $z_{\rm h}$ [pc] &  & $669_{-101}^{+116}$ & $161_{-30}^{+34}$ \\
     \vspace{1.2mm}
     & $\rho_{\rm dm}$ [${\rm M}_{\odot}\,{\rm pc}^{-3}$] &  & $0.0192_{-0.0023}^{+0.0023}$ & $0.0056_{-0.0033}^{+0.0039}$ \\
     \hline
     \vspace{1.2mm}
     \multirow{3}{*}{ Enlarge Error } & $\Sigma_{\star}$ [${\rm M}_{\odot}\,{\rm pc}^{-2}$] & $36.3_{-5.3}^{+5.3}$ & $37.4_{-5.2}^{+5.2}$ & $36.0_{-5.2}^{+5.0}$ \\
     \vspace{1.2mm}
     & $z_{\rm h}$ [pc] & $328_{-59}^{+59}$ & $441_{-77}^{+89}$ & $206_{-51}^{+59}$ \\
     \vspace{1.2mm}
     & $\rho_{\rm dm}$ [${\rm M}_{\odot}\,{\rm pc}^{-3}$] & $0.0119_{-0.0024}^{+0.0025}$ & $0.0135_{-0.0023}^{+0.0024}$ & $0.0077_{-0.0037}^{+0.0038}$ \\
     \hline
  \end{tabular}
\end{table*}
%%%%%%%%

%%%%%%%%%%% Fig of results with tilt term
\begin{figure}
   \centering
   \includegraphics[width=0.9\columnwidth]{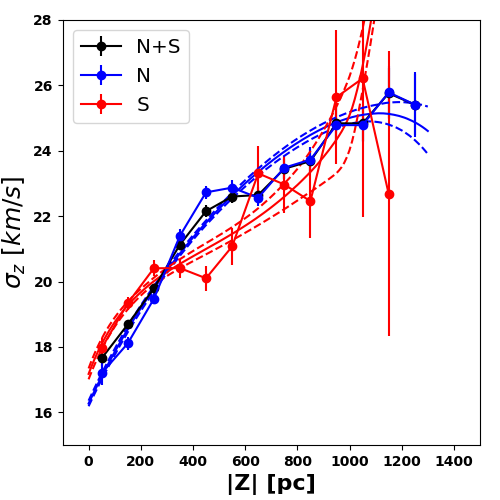}
   \caption{The model predicted $\sigma_{z}$ profiles of the northern (red lines) and southern (blue lines) subsamples with the asymmetric tilt term taken into consideration. The observed $\sigma_{z}$ profiles, i.e. the coloured dots, are the same as the right panel of Fig. \ref{fig_NS_profiles}. The dashed lines are the corresponding 1$\sigma$ errors.}
   \label{fig_NS_tilt}
\end{figure}
%%%%%%%%%%%

With the tilt term taken into consideration, we apply the Gaussian prior $P(\Sigma_{\star}) \sim {\rm N}(37.0,5.3^{2})\ {\rm M}_{\odot}\,{\rm pc}^{-2}$ to estimate the parameters. The model predicted parameters for the northern and southern subsamples are listed in Table \ref{tab_tilt_Lerr}. The model predicted $\sigma_{z}$ profiles are shown in Fig. \ref{fig_NS_tilt}. Both the northern and southern subsamples can be well fitted when they are considered separately. However, the tilt term gives a negative contribution to $\sigma_{z}$ for the northern subsample and a positive contribution to $\sigma_{z}$ for the southern subsample. Due to the different contributions of the tilt term, the model can not fit the northern and southern $\sigma_{z}$ profiles well at the same time.

The model predicted local dark matter densities for the northern and southern subsamples are $0.0192_{-0.0023}^{+0.0023}\ {\rm M}_{\odot}\,{\rm pc}^{-3}$ and $0.0056_{-0.0033}^{+0.0039}\ {\rm M}_{\odot}\,{\rm pc}^{-3}$, respectively. These values are similar to those obtained when the tilt term is ignored, as shown in Fig. \ref{fig_rhodm_phi}. The local dark matter densities of the northern and southern subsamples are still inconsistent. The density scale height of the southern subsample is much smaller than the usual 300 pc. Even when the tilt term is considered, the revised model can not explain the 2.6 ${\rm km\,s^{-1}}$ difference of $\sigma_{z}$ at $z = 450$ pc. The tilt term has small influence on the parameter estimations for the region considered, but it is not the primary reason for the asymmetry between the north and south.

\subsubsection{Dis-equilibrium}
\label{sssec_NS_Lerr}
Substantive observational evidence of vertical oscillations of the stellar disc has been found from photometrical and kinematical studies with different surveys \citep[e.g.][]{widrow2012, williams2013, xu2015, carrillo2018, wang2018, bennett2019, wang2019, gardner2020}. The Galaxy seems to have a ringing, wobbling, flaring and warped disc, which would cause deviations from mirror symmetry with respect to its mid-plane. These effects could be caused by some unknown systematics, for example dis-equilibrium. This could be due to the presence of a bar and spiral arms in the Milky Way \citep{bissantz2002, antoja2011, monari2016}, the `moving groups' in the Solar neighbourhood \citep{dehnen1998b}, or the vertical waves in the disc \citep{widrow2012, williams2013} caused by the Sagittarius merger and satellite perturbations \citep{purcell2011}.

Studies using stellar counts have confirmed the asymmetry of the Galactic disc \citep{widrow2012, yanny2013, bennett2019}. \cite{bennett2019} recently confirmed the density asymmetry with a maximum amplitude of $\sim$ 10\% for the main-sequence stars. However, the accurate density fluctuations of all types of stars are difficult to determine. The kinematic asymmetry between the north and south has been studied by LAMOST and Gaia \citep{wang2018, wang2019, bennett2019}. The vertical mean velocity and velocity dispersion profiles of our subsamples also show signs of north-south asymmetry, as shown in Fig. \ref{fig_NS_profiles}. 

Note that if the potential is asymmetric between the north and south, the calculation of $K_{z}$ force in Eq. \ref{eq_kz_int} will be problematic. In Eq. \ref{eq_kz_int}, a default assumption is applied: $\frac{{\rm d} \Phi}{{\rm d} z}|_{z=0} = 0$, i.e. the northern and southern mass profiles are symmetric. If this assumption is broken, the integration of $K_{z}$ force should be from a vertical height with $\frac{{\rm d} \Phi}{{\rm d} z} = 0$, rather than from zero to $z$ as previously. Overall, regardless of quite how gas or the stellar disc are modelled, accurate asymmetric mass models for the Galactic north and south are now difficult to obtain from observations. The application of such models is beyond our scope of this paper.

\begin{figure}
   \centering
   \includegraphics[width=0.9\columnwidth]{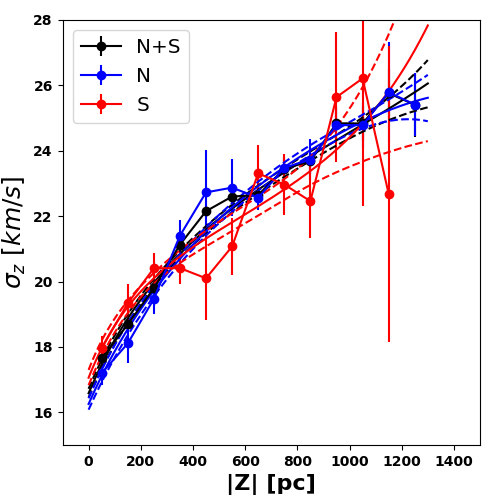}
   \caption{Same as Fig. \ref{fig_NS_tilt} but for the results when the data are binned and errors are enlarged.}
   \label{fig_NS_Lerr}
\end{figure}

In order to obtain consistent results for the northern and southern subsamples, we use half the difference between the $\sigma_{z}$ profiles as the measurement error, to give a simple alleviation of the asymmetry. Thus, the data need to be binned and we use the same bins as previous analyses. As the tilt term has small influence to the model, we ignore it here. The Gaussian prior on $\Sigma_{\star}$ in Table \ref{tab_prior} is used. The model predicted parameters for the data with enlarged errors are given in Table \ref{tab_tilt_Lerr}. The predicted $\sigma_{z}$ profiles are shown in Fig. \ref{fig_NS_Lerr}. The estimated local dark matter densities are $0.0119_{-0.0024}^{+0.0025}$ , $0.0135_{-0.0023}^{+0.0024}$ and $0.0077_{-0.0037}^{+0.0038}$ ${\rm M}_{\odot}\,{\rm pc}^{-3}$, for the total, the northern and the southern subsamples respectively. Although $z_{\rm h}$ of the southern subsample still seems to be underestimated, $\rho_{\rm dm}$ of the north and south are now consistent with each other (due to the enlarged errors).

%%%%%%%%%%% Fig of results with tilt term
%%%%%%%%%%%

%%%%%%%%%%%%%%%
\subsection{Mock tests for systematical uncertainties}
\label{ssec_error}
The original idea behind this work was to determine the local dark matter density with a higher accuracy when applying non-informative priors on all the model parameters. However, due to the data spatial distribution and the strong degeneracy between the dark matter and the stellar disc, the non-informative priors lead to a $\rho_{\rm dm}$ having a peak close to zero and large error bars. The non-informative priors obtain $\rho_{\rm dm} = 0.0049_{-0.0037}^{+0.0061}\ {\rm M}_{\odot}\,{\rm pc}^{-3}$, which has quite similar PDFs of model parameters as Fig. 11 of \cite{xia2016}. In \cite{xia2016}, they use mock data containing stars within $0 < z < 1000$ pc and apply non-informative priors on all parameters. They obtain a value of $\rho_{\rm dm}$ biased to zero, similar to our result of non-informative priors. They conclude that the mock data lacks stars with high $z$, where the velocity dispersion is more dominated by the contribution from the constant dark matter density. In addition, they argue that the Poisson noise from the sample size contributes about two-thirds of the uncertainty in the estimated values. An improvement in the sample size will increase the accuracy of the estimates. However, although our sample size is about two orders of magnitude larger than \cite{xia2016}, the non-informative priors still give quite large error bars.

\begin{figure}
   \centering
   \includegraphics[width=0.9\columnwidth]{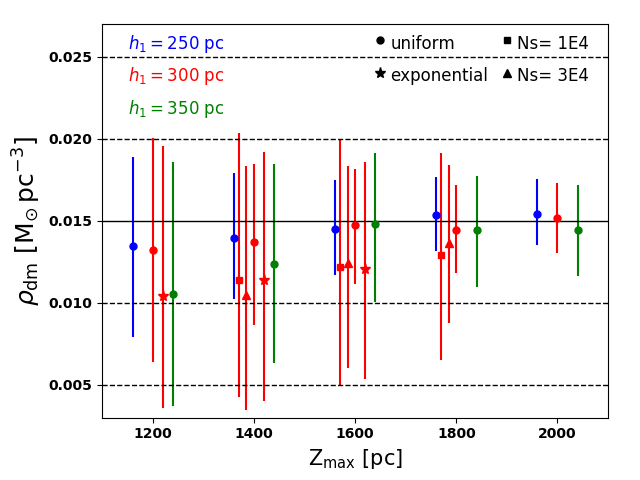}
   \caption{Error sources in the vertical Jeans equation method. The $x$-axis indicates five maximum heights (${\rm Z}_{\rm max}$) of the mock data, while the lower limits are set as zero for all mocks. The scale heights of the tracer populations are labelled with different colours as blue for $h_{1} = 250$ pc, red for $h_{1} = 300$ pc and green for $h_{1} = 350$ pc. The red stars show the mock data using the number density profile of the tracer to sample stars, while all the dots use uniform distributions in the ranges of $0 < z < {\rm Z}_{\rm max}$ for sampling. The dots and stars all have a sample size of 100,000 stars, while the squares and triangles contain 10,000 and 30,000 stars in each mock data set, respectively. The true value of $\rho_{\rm dm}$ is indicated with solid horizontal line. The dashed lines show some referenced values.}
   \label{fig_errors}
\end{figure}

To understand the error sources in our measurements, we make groups of mock data to check the influence of the sample size, the vertical range, the vertical stellar distribution and the scale height of the tracer population. For the mass models, we use the same set of parameters with $\Sigma_{\star} = 40\ {\rm M}_{\odot}\,{\rm pc}^{-2}$, $z_{\rm h} = 400\ {\rm pc}$, and $\rho_{\rm dm} = 0.015\ {\rm M}_{\odot}\,{\rm pc}^{-3}$. For each group of mock data, we make 50 sets of data to reduce the numerical noise. The estimated $\rho_{\rm dm}$ and its errors are calculated from the median values of the 50 sets of data for each group. In all mock data sets, non-informative priors are applied in MCMC. Our results are shown in Fig. \ref{fig_errors}.

%%%%%%%%%%%%%% Fig of error sources
%%%%%%%%%%%%%%

For the sample size, the mock data shown as dots and stars in Fig. \ref{fig_errors} have a sample size of 100,000 stars, comparable to the size of our observational sample. For comparisons, the mock data shown as squares and triangles have smaller sample sizes, containing 10,000 and 30,000 stars, respectively. As shown in Fig. \ref{fig_errors}, the accuracy increases as the sample size increases. Mock data sets with small sample size usually slightly underestimate $\rho_{\rm dm}$, as shown by the red squares in Fig. \ref{fig_errors}. A larger sample size will alleviate this bias.

For the vertical distribution of the mock data, we try exponential and uniform distributions in a range of $0 < z < {\rm Z}_{\rm max}$, where ${\rm Z}_{\rm max}$ is the maximum height of the mock data. We use the number density function $\nu_{z}$ to sample the exponentially distributed mock data, indicated by the red stars in Fig. \ref{fig_errors}. For comparisons, the mock data with a uniform distribution are shown as the red dots. The mock data with an exponential distribution obviously underestimates the dark matter density and has larger error bars. This is due to the non-binned MCMC, in which the low-$z$ regime dominates the likelihood calculation under an exponential distribution. As a result, the dark matter density will be underestimated. In observations, the sample of \cite{zhang2013} and our sample have column-like volumes, while the samples of \cite{garbari2012} and \cite{xia2016} have conical volumes. The former will have observational stellar distributions close to an exponentially distributed mock data. The latter sample stars with a weight of $z^{2}\nu_{z}$, and thus the high-$z$ regime has a larger contribution to the likelihood. Under the non-informative priors, \cite{zhang2013} and ourselves do obtain smaller $\rho_{\rm dm}$ than \cite{xia2016}. Relative to the MCMC, binning the data is similar to adding a weight, which changes the observational stellar distribution to a uniform distribution. This weighting is helpful to highlight the contribution of the dark matter. However, it will also magnify the numerical noise at high latitude, where the uncertainty in velocity is more significant.

In our mock tests, the vertical range of the data is the most prominent factor that influences the accuracy of the measurement. The mock data with a vertical range of ${\rm Z}_{\rm max} = 1200$ pc, i.e. $0 < z < 1200$ pc, have an error about 0.007 ${\rm M}_{\odot}\,{\rm pc}^{-3}$. Nevertheless, the mock data with $0 < z < 2000$ pc have an uncertainty only about 0.002 ${\rm M}_{\odot}\,{\rm pc}^{-3}$, even under the non-informative priors. A larger vertical range can more clearly separate the contributions from the exponential stellar disc and the constant dark matter density.

Another reason that a larger vertical range can increase the accuracy is indicated by the second term on the right-hand side of Eq. \ref{eq_sigz}. This term has a factor of $\exp \left( \frac{z}{h_{1}} \right)$, which contributes an exponential increase to the velocity dispersion. When $Z_{\rm max} = 2000$ pc and $h_{1} = 300$ pc, the factor is about 1000. Thus, in the high-$z$ regime, this term will try to match the observational velocity dispersion by tuning $f(z_{0})$ and thus the model parameters. Similarly, the scale height $h_{1}$ of the tracer population will also influence the measurement accuracy. As shown as coloured dots in Fig. \ref{fig_errors}, mock data with a smaller $h_{1}$ have smaller error bars. This is more prominent for data with a smaller vertical range.

For our observational sample, we have a vertical range of $0 < z < 1300$ pc and a scale height of 278.6 pc for the tracer population. However, due to the observational stellar vertical distribution, $\rho_{\rm dm}$ from the non-informative priors is biased toward a low value. In practical observations, tracers with smaller scale height usually have smaller vertical range. In our preliminary analysis of the K-giant stars, they are complete for $0 < z < 2000$ pc and have a scale height $\sim 360$ pc. An independant analysis of this K-giant sample will be helpful to constrain the range of $\rho_{\rm dm}$. We leave this to a future work.

Besides the observational limitations, the mass model also influences the uncertainty of the measurement. A larger $\rho_{\rm dm}$ can be more easily separated from the exponential disc. A smaller thin disc scale height $z_{\rm h}$ can make the contribution from the dark matter more significant in the high-$z$ regime. In addition, the ignored thick disc is thought to have a small contribution to the stellar disc and large scale height. Thus it has a contribution to the $K_{z}$ force similar to the constant dark matter in a range of $0 < z < 1300$ pc. Ignoring the thick disc may slightly overestimate the local dark matter density. We discuss this next.

%%%%%%%%%%%%%%%
\subsection{The thick disc}
\label{ssec_double}
As discussed in the eighth assumption in Section \ref{ssec_jeans}, the thick disc contributes $\sim$ 10\% to the total surface density at $z = 1300$ pc. The contribution of the ignored thick disc is distributed across the thin disc and dark matter. This will lead to a maximum uncertainty of 20\% in the estimation of the local dark matter density, using $\rho_{\rm dm} = 0.013 \pm 0.003\, {\rm M}_{\odot}\,{\rm pc}^{-3}$ (taken from \cite{mckee2015}). This uncertainty is also influenced by the real local dark matter density and the degeneracy between the single exponential disc model and the double disc model. In this section, we discuss the effect of including a thick disc in our models.

Taking the thick disc into consideration, the mass model becomes
%%%%%%%%
\begin{align}
  \rho_{\rm tot}(z) =& \rho_{\rm thin,0}\,{\rm exp} \left( -\frac{z}{z_{\rm h,thin}} \right) \nonumber\\ 
  &+\ \rho_{\rm thick,0}\,{\rm exp} \left( - \frac{z}{z_{\rm h,thick}} \right)\ +\ \Sigma_{\rm gas}\,\delta(z)\ +\ \rho_{\rm dm}\ ,
\label{eq_rho_dd}
\end{align}
%%%%%%%%
where $\rho_{\rm thin,0}$ and $\rho_{\rm thick,0}$ are the stellar volume densities on the Galactic plane of the thin and thick discs, respectively. $z_{\rm h,thin}$ and $z_{\rm h,thick}$ are the scale heights of the thin and thick discs, respectively. The scale heights are constrained to $0 < z_{\rm h,thin} < 800$ pc and $z_{\rm h,thin} < z_{\rm h,thick} < 1500$ pc in MCMC. Considering the strong degeneracy between the double discs and the dark matter, we apply both the total stellar surface density prior and the local stellar volume density prior listed in Table \ref{tab_prior}. Thus, $P(\rho_{\rm thin,0} + \rho_{\rm thick,0}) \sim {\rm N}(0.0468,0.0050^{2})\ {\rm M}_{\odot}\,{\rm pc}^{-3}$ and $P(\Sigma_{\star, {\rm thin}} + \Sigma_{\star, {\rm thick}}) \sim {\rm N}(37.0,5.3^{2})\ {\rm M}_{\odot}\,{\rm pc}^{-2}$, where $\Sigma_{\star} = 2\rho_{\star,0}z_{\rm h}$.

\begin{figure*}
   \centering
   \includegraphics[width=0.9\textwidth]{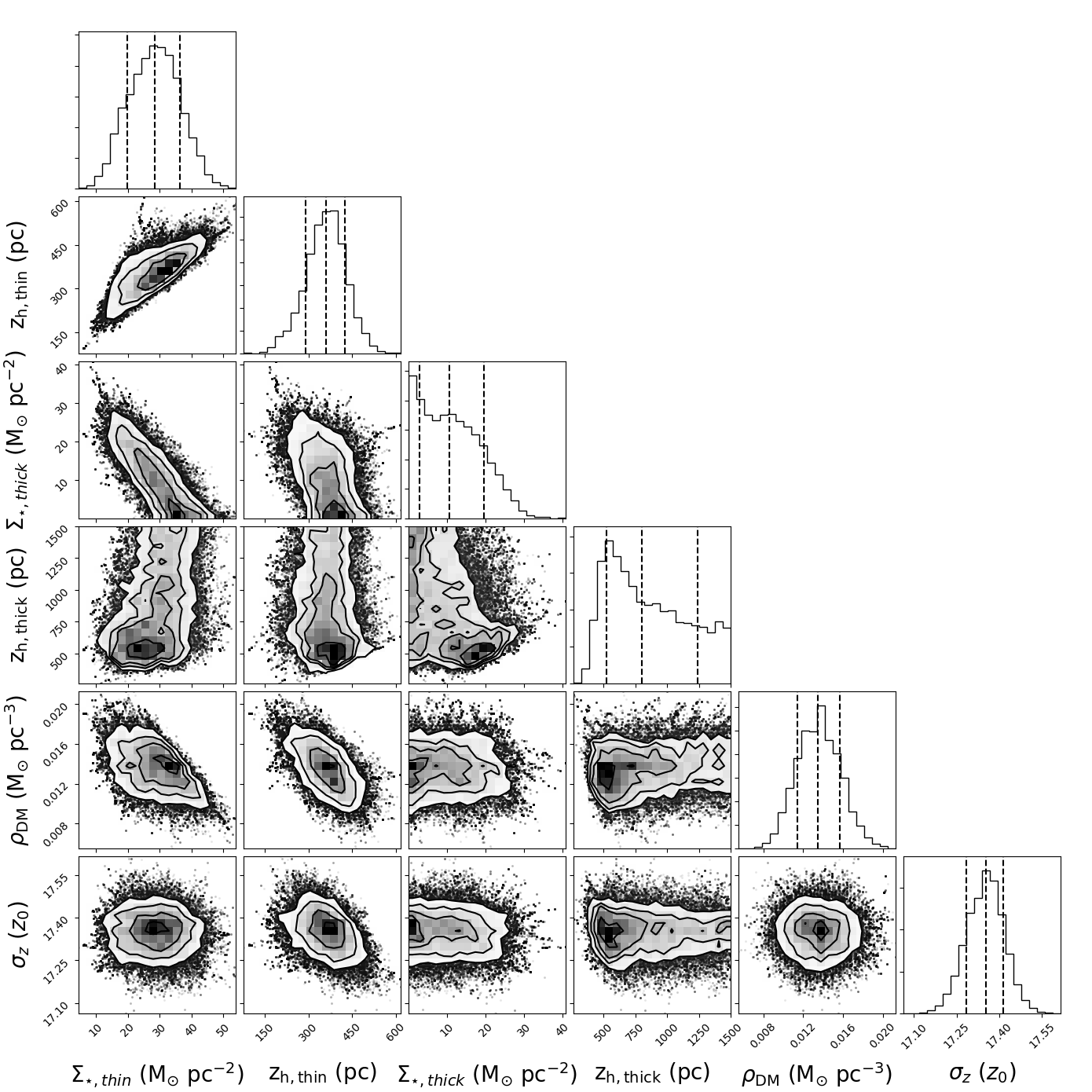}
   \caption{Similar to Fig. \ref{fig_tri} but for the double disc model.}
   \label{fig_dd_tri}
\end{figure*}

The parameters of the double disc model are $\boldsymbol{p} = (\Sigma_{\star, {\rm thin}},\ z_{\rm h,thin},\ \Sigma_{\star, {\rm thick}},\ z_{\rm h,thick},\ \rho_{\rm dm},\ \sigma_{z}(z_{0}))$. The PDFs of the model parameters are shown in Fig. \ref{fig_dd_tri}. The model predicted 1D marginalized parameters are ($28.5_{-8.6}^{+7.8}\ {\rm M}_{\odot}\,{\rm pc}^{-2}$, $362_{-70}^{+64}$ pc, $10.6_{-7.8}^{+9.0}\ {\rm M}_{\odot}\,{\rm pc}^{-2}$, $799_{-274}^{+443}$ pc, $0.0135_{-0.0021}^{+0.0022}\ {\rm M}_{\odot}\,{\rm pc}^{-3}$, $17.4_{-0.1}^{+0.1}\ {\rm km\,s^{-1}}$). The predicted midplane thick-to-thin disc density ratio $f = \rho_{\rm thick,0}/\rho_{\rm thin,0}$ is $0.153_{-0.12}^{+0.29}$. The median value of the ratio $f$ is similar to the 12\% given by \cite{juric2008} and consistent with \cite{siegel2002} ($\geq 10\%$). The model predicted $\rho_{\rm dm}$ is consistent with the result of the thin disc model.

However, $\Sigma_{\star, {\rm thick}}$ shows two peaks: one is at zero, which means that the thick disc is difficult to be recognized; another is near the median value of its 1D marginalized PDF. The second peak may be related to the best parameters with the maximum likelihood, which are (46.2 ${\rm M}_{\odot}\,{\rm pc}^{-2}$, 538 pc, 13.5 ${\rm M}_{\odot}\,{\rm pc}^{-2}$, 795 pc, 0.0062 ${\rm M}_{\odot}\,{\rm pc}^{-3}$, 17.3 ${\rm km\,s^{-1}}$). This set of parameters has a midplane thick-to-thin disc density ratio $f$ of 0.198 and a local dark matter density $\rho_{\rm dm}$ of 0.0062 ${\rm M}_{\odot}\,{\rm pc}^{-3}$. The local dark matter density is much smaller than that of the thin disc model.

We need to note that the double disc model is based on two strong priors listed in Table \ref{tab_prior}. It shows larger uncertainties, especially for $\Sigma_{\star, {\rm thick}}$ and $z_{\rm h,thick}$. The former has an uncertainty about 80\%, and the latter has uncertainty about 50\%. We also try to apply the three different priors individually to the double disc model. The model predicted local dark matter densities are $0.0134^{+0.0022}_{-0.0022}\, {\rm M}_{\odot}\,{\rm pc}^{-3}$, $0.0066^{+0.0046}_{-0.0038}\, {\rm M}_{\odot}\,{\rm pc}^{-3}$ and $0.0047^{+0.0054}_{-0.0032}\, {\rm M}_{\odot}\,{\rm pc}^{-3}$, for the Gaussian prior on $\Sigma_{\star}$, the Gaussian prior on $\rho_{\star,0}$ and the non-informative prior, respectively. These values are consistent with those from the single exponential disc model, shown in Table \ref{tab_para}. In addition, according to our mock data tests in the previous section, it is difficult to separate the contributions from the thick disc with that from the thin disc and the dark matter. A larger data vertical range would be helpful. We leave this to a future work with a stellar giants sample, in which the tilt term will also be carefully considered.

%%%%%%% figure of PDFs of double discs
%%%%%%%

\subsection{The gas component}
\label{ssec_gas}
In all our previous analyses, the total surface density of the gas is set as a constant. This assumption is also applied in \cite{bovy2013}, \cite{zhang2013} and \cite{xia2016}. However, measurements of gas usually have large uncertainties, both for the gas surface density and its scale height \citep{read2014, mckee2015, sivertsson2018}. The thickness of the HI disc is about 150 pc according to \cite{kalberla2009}. More recent measurements claim a scale height of 127 pc for the cold neutral medium and a scale height of 300-400 pc for the warm neutral medium \citep{mckee2015}. The effective scale height is about 200 pc according to \cite{mckee2015}. The influence of the gaseous thickness may not be negligible for the stars close to the Galactic plane.

To assess the influence of gaseous uncertainties, we try setting the total surface density and scale height of gas as free parameters. For the gas surface density, we apply a Gaussian prior or a non-informative prior to it. For the former, we apply $P(\Sigma_{\rm gas}) \sim {\rm N}(13.65, 2.78^{2})\ {\rm M}_{\odot}\,{\rm pc}^{-2}$ to the gas surface density \citep{sivertsson2018}. The predicted local dark matter density under the Gaussian prior on $\Sigma_{\star}$ is $0.0131 \pm 0.0019\ {\rm M}_{\odot}\,{\rm pc}^{-3}$, which is consistent with the result when $\Sigma_{\rm gas}$ is set as 13.2 ${\rm M}_{\odot}\,{\rm pc}^{-2}$. The results of these two gas models under the non-informative priors on other parameters are also consistent. For the non-informative prior on $\Sigma_{\rm gas}$, we merely constrain $5< \Sigma_{\rm gas} <22$ ${\rm M}_{\odot}\,{\rm pc}^{-2}$. The resultant local dark matter density is $0.0133 \pm 0.0022\ {\rm M}_{\odot}\,{\rm pc}^{-3}$, which is again consistent with the previous result. The predicted gas surface density is $14.9 \pm 3.2\ {\rm M}_{\odot}\,{\rm pc}^{-2}$, which is slightly larger than the quoted Gaussian prior on $\Sigma_{\rm gas}$. Considering the large $\Sigma_{\rm gas}$ uncertainties, these estimates of $\Sigma_{\rm gas}$ are consistent with each other.

For the gaseous thickness $Z_{\rm h,gas}$, we try fixing it as 150 pc or constraining it between 0 and 500 pc. The Gaussian prior $P(\Sigma_{\rm gas}) \sim {\rm N}(13.65, 2.78^{2})\ {\rm M}_{\odot}\,{\rm pc}^{-2}$ is applied for the gas total surface density for both situations. The Gaussian prior on $\Sigma_{\star}$ is also applied for the stellar thin disc. For the first situation ($Z_{\rm h,gas} = 150$ pc), the model predicted gaseous total surface density is $14.2 \pm 2.8\ {\rm M}_{\odot}\,{\rm pc}^{-2}$. The estimated local dark matter density is $0.0116^{+0.0028}_{-0.0026}\ {\rm M}_{\odot}\,{\rm pc}^{-3}$, which is slightly smaller but still consistent with the value when $\Sigma_{\rm gas} = 13.2\ {\rm M}_{\odot}\,{\rm pc}^{-2}$. For the second situation, the model predicted $Z_{\rm h,gas}$ is nearly zero, which means the model prefers a razor thin gaseous disc. The predicted local dark matter density of $0.0131 \pm 0.0022\ {\rm M}_{\odot}\,{\rm pc}^{-3}$ is consistent with that for the razor thin gaseous disc model.

From the above analyses, the uncertainties in $\Sigma_{\rm gas}$ seem not to significantly influence the local dark matter density estimation. More complicated gas models have been obtained from the observational gaseous census, and utilized in \cite{garbari2012}, \cite{mckee2015} and \cite{sivertsson2018}. They adopt different density profiles for different gaseous components, rather than a total surface density. This kind of complicated gas model is beyond the scope of this paper, and may not significantly alter our results as indicated by our previous analyses.

%%%%%%%%%%%%%%
\section{Conclusions}
\label{sec_conclu}
We utilize the vertical Jeans equation and the kinematics of stars in the solar neighbourhood to measure the local dark matter density. The sample is selected from the combined data set of LAMOST DR5 and Gaia DR2. Gaia DR2 provides proper motions, parallax and distance with high accuracies, and LAMOST DR5 provides good measurements of stellar radial velocity, stellar effective temperature and metallicity. With these parameters, we derive the vertical velocities for stars in a large volume, with the selection effects and volume completeness carefully taken into account. The selected sample contains more than 90,000 stars with a vertical range of $0 < |z| < 1300$ pc and an azimuthal angle range of $|\phi| < 5^{\circ}$. This sample is a factor of $\sim 70$ larger than the previous sample of \cite{xia2016}.

The number density profile of the selected sample is well fitted with a single exponential function with a scale height of 278.6 pc. For the mass models, we assume a single exponential stellar disc, a razor thin gas disc and constant dark matter. The tilt term and the circular velocity term are initially ignored as a simplification. With the simplified vertical Jeans equation and the non-binned MCMC simulations, we compare the model predicted vertical velocity dispersion profile with the observed vertical velocities, and obtain estimates for the model parameters. The total stellar surface density $\Sigma_{\star}$, the scale height $z_{\rm h}$ of the stellar disc and the dark matter density $\rho_{\rm dm}$ show strong degeneracy. Under a Gaussian prior on $\Sigma_{\star}$, compiled from previous literatures, the estimated $\rho_{\rm dm}$ is $0.0133_{-0.0022}^{+0.0024}\ {\rm M}_{\odot}\,{\rm pc}^{-3}$, and the predicted total surface density up to 1 kpc is $74.7_{-1.4}^{+1.4}\ {\rm M}_{\odot}\,{\rm pc}^{-2}$. These measurements are consistent with several previous works. However, using a Gaussian prior to the midplane stellar volume density and the non-informative priors give much lower measurements of $\rho_{\rm dm}$ ($0.0071_{-0.0043}^{+0.0059}$ and $0.0049_{-0.0037}^{+0.0061}$ ${\rm M}_{\odot}\,{\rm pc}^{-3}$, respectively).

We separate our sample into different azimuthal angle ranges and into northern and southern subsamples. The subsamples with different azimuthal angle ranges have similar $\rho_{\rm dm}$ measurements. However, the velocity dispersion profiles of the northern and southern subsamples have plateaux in different vertical regions. The plateaux give strong constraints on the scale heights $z_{\rm h}$. Consequently, the estimated local dark matter densities show large discrepancies using the same prior. The tilt term is reconsidered and it has small contribution to $\sigma_{z}$ in the vertical region considered. Though the derivative of the tilt term with respect to R seems to be asymmetric in the north and south, it does not explain the $\sigma_{z}$ asymmetry we found. Taking half of the $\sigma_{z}$ difference between the northern and southern subsamples as errors to account for unknown systematics, we obtain consistent $\rho_{\rm dm}$ for the north and south. The model predicted $\rho_{\rm dm}$ for the total sample is then $0.0119_{-0.0024}^{+0.0025}\ {\rm M}_{\odot}\,{\rm pc}^{-3}$.

We make groups of mock data to examine the uncertainty in the determination of the local dark matter density. An increase in sample size can improve the measurement accuracy, but this improvement is limited by the spatial distribution of the sample. An exponentially distributed sample will highlight the contribution of stars in low-$z$ regime and underestimates the local dark matter density. This is similar to our results using non-informative priors. A large vertical range and a small scale height of the tracer population can more easily break the degeneracy between the model parameters, and thus decrease the uncertainty of the local dark matter density significantly. However, these two factors are usually correlated and restricted by observational conditions. A double disc model is also considered under strong priors. The parameters with the maximum likelihood give a midplane thick-to-thin disc density ratio of 0.198 and a local dark matter density of 0.0062 ${\rm M}_{\odot}\,{\rm pc}^{-3}$.

In future works, a stellar giants sample with good $\alpha$-element measurements and a larger vertical range ($0 - 2000$ pc) will be considered. Both independent analysis and combined analysis with the sample in this work will be helpful in determining the local dark matter density more accurately. In addition, the structures in the velocity dispersion profiles and the tilt term need to be carefully checked and better understood in the future.

\section*{Acknowledgements}
We are grateful to the anonymous referee for thoughtful comments that much improved the paper. We thank useful discussions with Professors Lia Athanassoula, Konrad Kuijken and James Binney. This work is partly supported by the National Key Basic Research and Development Program of China (No. 2018YFA0404501 to SM and No. 2019YFA0405501 to CL), by the National Science Foundation of China (Grant No. 11821303, 11761131004 and 11761141012 to SM and No. 11835057 to CL). This project was developed in part at the 2018 Gaia-LAMOST Sprint workshop, supported by the National Natural Science Foundation of China (NSFC) under grants 11333003 and 11390372. X. -X-X thanks the support by the National Key R\&D Program of China under grant No. 2019YFA0405500 and by NSFC grant No. 11988101, 11873052, 11890694.

%%%%%%%%%%%%%%%%%%%%%%%%%%%%%%%%%%%%%%%%%%%%%%%%%%



%\bibliographystyle{mnras}


%%%%%%%%%%%%%%%%%%%%%%%%%%%%%%%%%%%%%%%%%%%%%%%%%%

%%%%%%%%%%%%%%%%% APPENDICES %%%%%%%%%%%%%%%%%%%%%
%\appendix

%%%%%%%%%%%%%%%%%%%%%%%%%%%%%%%%%%%%%%%%%%%%%%%%%%

% Don't change these lines
\bsp	% typesetting comment
\label{lastpage}
\end{document}